\documentclass[11pt,A4 paper]{article}

\usepackage{colortbl}
\usepackage[margin = 1in]{geometry}
\usepackage{microtype}
\usepackage{epsfig}
\usepackage{microtype}
\usepackage{graphicx}
\usepackage{enumitem}
\usepackage{multirow}
\usepackage{epsfig,enumerate,amsmath,amsfonts,amssymb,amsthm,mathrsfs,ifpdf}
\usepackage{indentfirst,relsize}
\usepackage[numbers]{natbib}
\usepackage{setspace}

\usepackage{enumerate}
\usepackage{latexsym}
\usepackage{stackrel}
\usepackage[linesnumbered,vlined, ruled]{algorithm2e}
\usepackage[all]{xy}
\usepackage[usenames,dvipsnames]{pstricks}
\usepackage{pst-grad} 
\usepackage{pst-plot} 
\usepackage{xspace}

\newcommand{\size}[1]{\left| #1 \right|}

\newcommand{\remove}[1]{}

\newcommand{\R}{\mathbb{R}}
\newcommand{\N}{\mathbb{N}}

\newcommand{\cZ}{\mathcal{Z}}
\newcommand{\cM}{\mathcal{M}}

\newcommand{\cA}{\mathcal{A}}

\newcommand{\cC}{\mathcal{C}}

\newcommand{\cF}{\mathcal{F}}
\newcommand{\cP}{\mathcal{P}}

\newcommand{\Oh}{\mathcal{O}}

\newcommand{\pr}{\mathbb{P}}

\newcommand{\disj}{$\mbox{{\sc Disj}}_n$\xspace}
\newcommand{\disjxy}{\mbox{{\sc Disj}}_n({\bf x}, {\bf y})\xspace}
\newcommand{\ind}{$\mbox{{\sc Index}}_n$\xspace}
\newcommand{\indxy}{\mbox{{\sc Index}}_n({\bf x},j)\xspace}
\newcommand{\perm}{$\mbox{{\sc Perm}}_n$\xspace}
\newcommand{\permxy}{\mbox{{\sc Perm}}_n(\pi,j)\xspace}
\newcommand{\eam}{{\sc Edge Arrival}\xspace}
\newcommand{\ea}{\mbox{{\sc Ea}}\xspace}
\newcommand{\deam}{{\sc Dynamic Edge Arrival}\xspace}
\newcommand{\dea}{\mbox{{\sc Dea}}\xspace}
\newcommand{\vam}{{\sc Vertex Arrival}\xspace}
\newcommand{\va}{\mbox{{\sc Va}}\xspace}
\newcommand{\alm}{{\sc Adjacency List}\xspace}
\newcommand{\al}{\mbox{{\sc Al}}\xspace}
\newcommand{\vc}{\mbox{{\sc VC}}}

\newcommand{\bit}{\mbox{ bit}}

\newcommand{\bx}{{\bf x}}
\newcommand{\by}{{\bf y}}
\newcommand{\yes}{{\sc YES} \xspace}
\newcommand{\no}{{\sc NO} \xspace}

\newcommand{\cn}{{\sc Common Neighbor}\xspace}
\newcommand{\cvdsol}{\mbox{{\sc cvd}} \xspace}
\newcommand{\fvs}{{\sc FVS}\xspace}
\newcommand{\ect}{{\sc ECT}\xspace}
\newcommand{\Oct}{{\sc OCT}\xspace}
\newcommand{\tdel}{{\sc TD}\xspace}

\newcommand{\cvd}{{\sc CVD}\xspace}
\newcommand{\subdel}{{\sc $\cF$-Subgraph deletion}\xspace}
\newcommand{\minordel}{{\sc $\cF$-Minor deletion}\xspace}

\theoremstyle{plain}
\newtheorem{theo}{Theorem}[section]
\newtheorem{lem}[theo]{Lemma}
\newtheorem{pre}[theo]{Proposition}
\newtheorem{coro}[theo]{Corollary}

\newtheorem{cl}[theo]{Claim}
\theoremstyle{definition}
\newtheorem{defi}[theo]{Definition}
\newtheorem{rem}{Remark}
\newtheorem{obs}[theo]{Observation}

\newcommand{\defproblem}[3]{
  \vspace{1mm}
\noindent\fbox{
  \begin{minipage}{0.96\textwidth}
  \begin{tabular*}{\textwidth}{@{\extracolsep{\fill}}lr} #1 \\ \end{tabular*}
  {\bf{Input:}} #2  \\
  {\bf{Output:}} #3
  \end{minipage}
  }
  \vspace{1mm}
}

\usepackage{xcolor}
\usepackage{mdframed}

\title{
Structural Parameterization for Graph Deletion Problems \\over Data Streams~\footnote{A preliminary version of this article has been submitted to ITCS'2020.}
}
\date{}

\author{
Arijit Bishnu
\footnote{
Indian Statistical Institute, Kolkata, India
}
\and
Arijit Ghosh
\footnotemark[1]
\and
Sudeshna Kolay
\footnote{Ben-Gurion University of the Negev, Israel}
\and
Gopinath Mishra
\footnotemark[1]
\and 
Saket Saurabh
\footnote{The Institute of Mathematical Sciences, HBNI, India}
}
\begin{document}

\maketitle

\thispagestyle{empty}

\begin{abstract}

The study of parameterized streaming complexity on graph problems was initiated by Fafianie et al. (MFCS'14) and Chitnis et al. (SODA'15 and SODA'16). Simply put, the main goal is to design streaming algorithms for parameterized problems such that $\Oh(f(k) \log ^{O(1)} n)$ space is enough, where $f$ is an arbitrary computable function depending only on the parameter $k$. However, in the past few years very few positive results have been established. Most of the graph problems that do have streaming algorithms of the above nature are ones where localized checking is required, like {\sc Vertex Cover} or {\sc Maximum Matching} parameterized by the size $k$ of the solution we are seeking. Many important parameterized problems that form the backbone of traditional parameterized complexity are known to require $\Omega(n)$ bits for any streaming algorithm; e.g. {\sc Feedback Vertex Set}, {\sc Even Cycle Transversal}, {\sc Odd Cycle Transversal}, {\sc Triangle Deletion} or the more general {\sc ${\cal F}$-Subgraph Deletion} when parameterized by solution size $k$. 

Our main conceptual contribution is to overcome the obstacles to efficient parameterized streaming algorithms by utilizing the power of parameterization. To the best of our knowledge, this is the first work in parameterized streaming complexity that considers structural parameters instead of the solution size as parameter. We focus on the vertex cover size $K$ as the parameter for the parameterized graph deletion problems we consider. At the same time, most of the previous work in parameterized streaming complexity was restricted to the {\sc Ea} (edge arrival) or {\sc Dea} (dynamic edge arrival) models. In this work, we consider the four most well-studied streaming models: the {\sc Ea}, {\sc Dea}, {\sc Va} (vertex arrival) and {\sc Al} (adjacency list) models. Surprisingly, the consideration of vertex cover size $K$ in the different models leads to a classification of positive and negative results for problems like {\sc ${\cal F}$-Subgraph Deletion}, {\sc Cluster Vertex Deletion}{} (\cvd{})) and {\sc ${\cal F}$-Minor Deletion}. 

\remove{Our study also encompasses a set of lower bound results. We show that except for the {\sc Triangle Deletion}{} and {\sc Cluster Vertex Deletion}{} problems, none of the other problems have space-efficient streaming algorithms when the problems are parameterized by $k$, the solution size. Interestingly, the lower bounds do not depend on $k$; they depend only on $n$. All the other problems like {\sc ${\cal F}$-Minor Deletion}{} and {\sc ${\cal F}$-Subgraph Deletion} and its many special variants like {\sc Feedback Vertex Set}{}, {\sc Even Cycle Transversal}{}, {\sc Odd Cycle Transversal}{} admit $\Omega(n\log n)$ one pass lower bounds in all the four models stated above. This improves on the the one pass lower bounds given by Chitnis et al.(SODA'16) for the {\sc Ea} model in SODA'16.  
We show other lower bound results that are parameterized by the vertex cover size $K$ in other streaming models like \ea and \va. All these results extend the results of Chitnis et al. (SODA'16). To the best of our knowledge, this is the first set of lower bound results in the {\sc Al} model. 
}
\end{abstract}

\newpage
\pagestyle{plain}
\setcounter{page}{1}
\section{Introduction}
\label{sec:intro}
\noindent
In streaming algorithms, a graph is presented as a sequence of edges. In the simplest of this model, we have
a stream of edge arrivals, where each edge adds to the graph seen so far, or may
include a dynamic mixture of arrivals and departures of edges. In either case, the primary objective is to quickly answer some basic questions over the current state of the graph, such as finding a (maximal) matching over the current graph edges, or finding a (minimum) vertex cover, while storing only a small amount of information.  In the most restrictive model, we only allow $\Oh( \log ^{O(1)} n)$ bits of space for storage. However, using methods from communication complexity  one can show that most problems do not admit such algorithms. Thus one relaxes this notion and  defines what is 
called a {\em semi-streaming model}, which allows  $\Oh(n \log ^{O(1)} n)$  bits of space.  This model is extremely successful for graph streaming algorithms and a plethora of non-trivial algorithms has been made in this model~\cite{AssadiKL16,GuruswamiVV17,KapralovKSV17}.  There is a vast literature on graph streaming and we refer to the survey by McGregor~\cite{McGregor14} for more details. 

The theme of this paper is parameterized streaming algorithms. So, before we go into parametrized 
streaming let us introduce a few basic definitions in parameterized complexity. The goal of parameterized complexity is to find ways of solving NP-hard problems more efficiently than brute force: the aim is to
restrict the combinatorial explosion to a parameter that is hopefully
much smaller than the input size. Formally, a {\em parameterization}
of a problem is assigning an integer $k$ to each input instance. A parameterized problem is said to be {\em fixed-parameter tractable (FPT)} if there is an algorithm that solves the problem in time
$f(k)\cdot |I|^{O(1)}$, where $|I|$ is the size of the input and $f$ is an
arbitrary computable function depending only on the parameter $k$. There is a long list of NP-hard graph problems that are FPT under
various parameterizations: finding a vertex cover of size $k$, finding
a cycle of length $k$, finding a maximum independent set in a graph
of treewidth at most $k$, etc.
For more details, the reader is referred to the monograph~\cite{saketbook15}. 
Given the definition of FPT for parameterized problems, it is natural to expect an efficient algorithm for the corresponding parameterized streaming versions to allow $\Oh(f(k) \log ^{O(1)} n)$  bits of space, where $f$ is an
arbitrary computable function depending on the parameter $k$.

There are several ways to formalize the parameterized streaming question, and in literature certain natural models are considered. The basic case is when the input of a given problem consists of a sequence of edge arrivals only, for which one seeks a parameterized streaming algorithm (PSA). It is more challenging when the input stream is dynamic, and  contains both deletions and insertions of edges. In this case one seeks a dynamic parameterized streaming algorithm (DPSA). Notice that when an edge in the matching is deleted, we sometimes need substantial work to repair the solution and have to ensure that the algorithm has enough information to do so, while keeping only a bounded amount of working space. If we are promised that at every timestamp there is a solution of cost $k$, then we seek a promised dynamic parameterized streaming algorithm (PDPSA). These notions were formalized in the following two papers~\cite{ChitnisCEHMMV16, ChitnisCEHM15} and several results for  {\sc Vertex Cover} and {\sc Maximum Matching} were presented there.  Unfortunately, this relaxation to $\Oh(f(k) \log ^{O(1)} n)$  bits of space does not buy us too many new results. Most of the problems for which parameterized streaming algorithms are known are ``local problems''. Other local problems like {\sc Cluster Vertex Deletion} and {\sc Triangle Deletion} do not have positive results. Also, problems that require some global checking -- such as {\sc Feedback Vertex Set}, {\sc Even Cycle Transversal}, {\sc Odd Cycle Transversal} etc. remain elusive.  In fact, one can show that, when edges of the graph arrive in an arbitrary order, using reductions from communication complexity all of the above problems will require $\Omega(n)$ space even if we allow a constant number of passes over the data stream~\cite{ChitnisCEHMMV16}.


The starting point of this paper is the above mentioned $\Omega(n)$ lower bounds on basic graph  problems. We ask the most natural question -- how do we deconstruct these intractability results? When we look deeper we realize that, to the best of our knowledge the only parameter that has been used in parameterized streaming algorithms is {\em the size of the solution that we are seeking}.  Indeed this is the most well-studied parameter, but there is no reason to only use solution size as a parameter. 
\begin{quote}
 \begin{mdframed}[backgroundcolor=yellow!25]  
In parameterized complexity, when faced with such obstacles, we either study a problem with respect to parameters larger than the solution size or consider some structural parameters. We
export this approach to parameterized streaming algorithms. This
is our main conceptual contribution, that is, to introduce the concept of structural parameterizations
to the study of parameterized streaming algorithms.
    \end{mdframed}
\end{quote}
\subsection*{Parameters, models, problems and our results}
\noindent
{\bf What parameters to use?}
 In parameterized complexity, after solution size and treewidth, arguably the most notable structural parameter is  {\em vertex cover size $K$}~\cite{saketbook15,FOMIN2014468}. For all the vertex deletion problems that we consider in this paper, a vertex cover is also a solution. Thus, the vertex cover size $K$ is always larger than the solution size $k$ for all the above problems. We do a thorough study of  vertex deletion problems from the view point of parameterized streaming in all known models and show dichotomy when moving across parameters and streaming models. The main \emph{conceptual} contribution of this paper is to introduce structural parameter in parameterized streaming algorithms. 
 
 \medskip 
\noindent
{\bf Streaming models} 
 The models that we consider are:  (1) {\eam(\ea) model}; (2)  
{\deam(\dea) model}; (3)   {\vam(\va) model} (In a step, a vertex $v \in V(G)$ is exposed along with all the edges between $v$ and  already exposed neighbors of $v$.); and (4) {\alm(\al) model} (In a step, a vertex $v \in V(G)$ is exposed along with all edges incident on $v$). The formal definitions are in Section~\ref{sec:prelim-main}. 

\medskip 
\noindent
{\bf What problems to study?}
We study the streaming complexity of parameterized versions of \subdel, \minordel and {\sc Cluster Vertex Deletion}{} 
(\cvd{}). These problems are one of the most well studied ones in parametertized complexity~\cite{DBLP:journals/ipl/Cai96,DBLP:journals/talg/CaoM15,DBLP:journals/algorithmica/CaoM16,FOMIN2014468,FominLMPS11,FominLMS12,LPRRSS13,KociumakaP14,DBLP:journals/algorithmica/Marx10,ReedSV04,Thomasse10} and have led to development 
of the field. The parameters we consider in this paper are (i) the solution size $k$ and (ii) the size $K$ of the vertex cover 
of the input graph $G$. In \subdel, \minordel and \cvd, the objective is to decide whether there exists $X \subset V(G)$ of 
size at most $k$ such that $G \setminus X$ has no graphs in $\cF$ as a subgraph, has no graphs in $\cF$ as a minor and has 
no induced $P_3$, respectively. \subdel, \minordel and \cvd are interesting due to the following reasons. {\sc Feedback 
Vertex set} (\fvs), {\sc Even Cycle Transversal} (\ect), {\sc Odd Cycle Transversal} (\Oct) and {\sc Triangle Deletion} 
(\tdel) are special cases of \subdel when $\cF =\{C_3,C_4,C_5,\ldots\}$, $\cF=\{ C_3,C_5,\ldots\} $, $\cF =\{C_4,C_6,\ldots
\}$ and $\cF=\{C_3\}$, respectively. \fvs is also a special case of \minordel when $\cF=\{C_3\}$. \cvd is different as we 
are looking for induced structures.

\medskip
\noindent
{\bf Our results.}
 Let a graph $G$ and a non-negative integer $k$ be the inputs to the graph problems we consider. 
Notice that for \subdel, \minordel and \cvd, $K \geq k$. Interestingly, the {\em parameter $K$ also has different effects on the above mentioned problems in the different streaming models}. We show that structural parameters help to obtain efficient parameterized streaming algorithms for some of the problems, while no such effect is observed for other problems. This throws up the more general and deeper question in parameterized streaming complexity of classification of problems based on the different graph streaming models and different parameterization. We believe that our results and concepts will be instrumental in opening up the  avenue for such studies in future.  

In particular, we obtain a range of streaming algorithms as well as lower bounds on streaming complexity for the problems we consider. Informally, for a streaming model $\mathcal{M}$ and a parameterized problem $\Pi$, if there is a $p$-pass randomized streaming algorithm for $\Pi$ that uses $\Oh(\ell)$ space then we say that $\Pi$ is $(\mathcal{M},\ell,p)$-streamable. Similarly, if there is no $p$-pass algorithm using $o(\ell)$ bits{\footnote{It is usual in streaming that the lower bound results are in bits, and the upper bound results are in words.} of storage then $\Pi$ is said to be $(\mathcal{M},\ell,p)$-hard. For formal definitions please refer to Section~\ref{sec:prelim-main}. When we omit $p$, it means we are considering one pass of the input stream. 
The highlight of our results are captured by the \subdel, \minordel and \cvd problems. 

\begin{theo}
\label{theo:main2}
Consider \subdel in the \al model. Parameterized by solution size $k$, \subdel is $(\al,\Omega(n \log n))$-hard. However, when parameterized by vertex cover $K$, \subdel is $(\al,\Oh\left(\Delta(\cF) \cdot K^{\Delta(\cF)+1} \right))$-streamable. Here $\Delta(\cF)$ is the maximum degree of any graph in $\cF$. 
\end{theo}

The above Theorem is in contrast to results shown in~\cite{ChitnisCEHMMV16}. First, we would like to point out that to the best of our knowledge this is the first set of results on hardness in the \al model. The results in ~\cite{ChitnisCEHMMV16} showed that \subdel is $(\ea,\Omega(n))$-hard. A hardness result in the \al model implies one in the \ea model (Refer to Section~\ref{sec:prelim-main}). Thus, our result (Proof in Theorem~\ref{theo:lowerbounds1}) implies a stronger lower bound for \subdel particularly in the \ea model. On the positive side, we show that \subdel parameterized by the vertex cover size $K$, is $\left(\al,\Delta(\cF) \cdot K^{\Delta(\cF)+1}\right)$-streamable (Proof in Theorem~\ref{theo:sub_ub}). 

Our hardness results are obtained from reductions from well-known problems in \emph{communication complexity}. The problems we reduced from are \ind, \disj and \perm (Please refer to Section~\ref{append-sec:prelim} for details). In order to obtain the algorithm, one of the main technical contributions of this paper is the introduction of the \cn problem  which plays a crucial role in designing streaming algorithms in this paper. We show that \subdel and many of the other considered problems, like \minordel parameterized by vertex cover size $K$, have a unifying structure that can be solved via \cn, when the edges of the graph are arriving in the \al model. In \cn, the objective is to obtain a subgraph $H$ of the input graph $G$ such that the subgraph contains a maximal matching $M$ of $G$. Also, for each pair of vertices $a,b \in V(M)$~\footnote{$V(M)$ denotes the set of all vertices present in the matching $M$}, the edge $(a,b)$ is present in $H$ if and only if $(a,b) \in E(G)$, and \emph{enough}~\footnote{By enough, we mean $\Oh(K)$ in this case.} common neighbors of all subsets of at most $\Delta(\cF)$ vertices of $V(M)$ are retained in $H$. Using structural properties of such a subgraph, called the \emph{common neighbor subgraph}, we show that it is enough to solve \subdel on the common neighbor subgraph. Similar algorithmic and lower bound results can be obtained for \minordel. The following theorem can be proven using Theorem~\ref{theo:minor_ub} in Section~\ref{sec:minor} and Theorem~\ref{theo:lowerbounds1} in Section~\ref{sec:lbproof}.

\begin{theo}
\label{theo:main3}
Consider \minordel in the \al model. Parameterized by solution size $k$, \minordel is $(\al,\Omega(n \log n))$-hard. However, when parameterized by vertex cover $K$, \minordel is $(\al,\Oh\left(\Delta(\cF) \cdot K^{\Delta(\cF)+1} \right))$-streamable. Here $\Delta(\cF)$ is the maximum degree of any graph in $\cF$. 
\end{theo}
 
The result on \cvd is stated in the following Theorem. 
\begin{theo}
\label{theo:main1}
Parameterized by solution size $k$, \cvd is $(\va,\Omega(n))$-hard. However, when parameterized by vertex cover $K$ \cvd is $(\dea,\Oh\left(K^2 \log ^4 n\right))$-streamable. 
\end{theo}

The \cvd problem behaves very differently from the above two problems. We show that the problem is $(\va,n)$-hard (Theorem~\ref{theo:lowerbounds3}). In contrast, in~\cite{ChitnisCEHMMV16} the $(\ea,n)$-hardness for the problem was shown, and we are able to extend this result to the \va model (Refer to Section~\ref{sec:prelim-main} for relations between the models considered). Surprisingly, when we parameterize by  $K$, \cvd is $(\dea,K^2\log^4 n)$-streamable (Theorem~\ref{theo:cvd_ub}). In fact, this implies $(\cM,K^2\log^4 n)$-streamability for $\cM \in \{\mbox{\al,\va,\ea}\}$. To design our algorithm, we build on the sampling technique for {\sc Vertex Cover}~\cite{ChitnisCEHMMV16} to solve \cvd in \dea model. Our analysis of the sampling technique exploits the structure of a cluster graph.

Though we have mentioned the main algorithmic and lower bound result in the above theorems, we have a list of other algorithmic and lower bound results in the different streaming models. The full list of results are summed up in Table~\ref{table:lb}. To understand the full strength of our contribution, we request the reader to go to Section~\ref{sec:prelim-main} to see the relations between different streaming models and the notion of \emph{hardness} and \emph{streamability.}

\begin{table*}[t]
\label{table}
\small
\begin{center}
\begin{tabular}{|c | c | c | c |c | c }

\hline 
Problem & Parameter & \al model & \va model & \ea/\dea model \\
 \hline \hline

 \hline
 
   \multirow{ 3}{*}{{$\cF$-{\sc Subgraph}}} & \multirow{2}{*}{$k$} & {$(\al, n \log n)$-hard} & $(\va, n \log n)$-hard & $(\ea, n \log n)$-hard \\
  &  & {$(\al, n/p,p)$-hard} & $(\va, n/p,p)$-hard & $(\ea, n/p,p)$-hard$^\dag$\\ \cline{2-5}
{\sc Deletion} &   &  &  &  \\      
                     
                       &  $K$ & {}{ $(\al,\Delta(\cF) \cdot K^{\Delta(\cF)+1})$-str.$^{*}$} & {$(\va, n/p,p)$-hard} & $(\ea, n/p,p)$-hard \\
                       & & (Theorem~\ref{theo:sub_ub}) & & \\
                         \hline
                         
   \multirow{ 3}{*}{{$\cF$-{\sc Minor}}} & \multirow{2}{*}{$k$} & {$(\al, n \log n)$-hard} & $(\va, n \log n)$-hard & $(\ea, n \log n)$-hard \\
  &  & {$(\al, n/p,p)$-hard} & $(\va, n/p,p)$-hard & $(\ea, n/p,p)$-hard\\ \cline{2-5}
        {\sc Deletion}  &   &  &  &  \\               
                       &  $K$ & {}{ $(\al,\Delta(\cF) \cdot K^{\Delta(\cF)+1})$-str.$^{*}$} & {$(\va, n/p,p)$-hard} & $(\ea, n/p,p)$-hard \\
                          & & (Theorem~\ref{theo:minor_ub}) & & \\
 \hline  
 
{\fvs}, & \multirow{2}{*}{$k$} &{} {$(\al, n \log n)$-hard} & $(\va, n \log n)$-hard & $(\ea, n \log n)$-hard \\ 
        {\ect},               &  &{} {$(\al, n/p,p)$-hard} & $(\va, n/p,p)$-hard & $(\ea, n/p,p)$-hard$^\dag$\\ \cline{2-5}
       {\Oct}                  & $K$ & {$(\al,K^3)$-str.$^{*}$} & {} {$(\va, n/p,p)$-hard} & $(\ea, n/p,p)$-hard \\
                  & & (Corollary~\ref{theo:cycle_ub}) & & \\      
             \hline

 \multirow{ 3}{*}{\tdel} &\multirow{2}{*}{$k$} & \multirow{2}{*}{OPEN} &{} { $(\va, n \log n)$-hard} & $(\ea, n \log n)$-hard \\ 
                       & &  & {} {$(\va, n/p,p)$-hard}  & $(\ea, n/p,p)$-hard$^\dag$  \\ \cline{2-5}
                       &  $K$ & { $(\al,K^3)$-str.$^{*}$} & {} {$(\va, n/p,p)$-hard} & $(\ea, n/p,p)$-hard \\
                        & & (Corollary~\ref{theo:cycle_ub}) & & \\
 \hline

 \multirow{3}{*}{\cvd} & {$k$} & {OPEN} & {} { $(\va, n/p,p)$-hard} &  $(\ea, n/p,p)$-hard$^\dag$ \\ \cline{2-5}
                       & $K$ & $(\al,K^2\log^4 n)$-str. & $(\va,K^2\log^4 n)$-str. & {} {$(\dea,K^2\log^4 n)$-str.} \\
           & & (Theorem~\ref{theo:cvd_ub}) & & \\              
 \hline

\end{tabular}

\end{center}
\caption{A summary of our results. ``str." means streamable. The results marked with $\dag$ in Table~\ref{table:lb} are lower bound results of Chitnis et al.~\cite{ChitnisCEHMMV16}. The other 
lower bound results are ours, some of them being improvements over the lower bound results of Chitnis et al.~\cite{ChitnisCEHMMV16}. The full set of lower bound results for \fvs, \ect, \Oct are proven in Theorem~\ref{theo:lowerbounds1}. The lower bound results for \tdel and \cvd are proven in Theorem~\ref{theo:lowerbounds2} and Theorem~\ref{theo:lowerbounds3}, respectively. Notice that the lower bound results depend only on $n$. The hardness results are even stronger than what is mentioned here; the nuances are mentioned in respective Theorems\remove{~\ref{theo:lowerbounds1},~\ref{theo:lowerbounds2},~\ref{theo:lowerbounds3} in Section~\ref{sec:lbproof}. Algorithmic results marked $*$ are deterministic.}}  
\label{table:lb}
\end{table*}




\smallskip 
\noindent
{\bf Related Work.}
Problems in class P have been extensively studied in streaming complexity in the last 
decade~\cite{McGregor14}. Recently, there has been a lot of interest in studying streaming complexity of NP-hard problems like {\sc Hitting Set}, 
{\sc Set Cover}, {\sc Max Cut} and  
{\sc Max CSP}~\cite{GuruswamiVV17,KapralovKSV17,AssadiKL16}.\remove{ Some notable results include 
Kapralov {\em et al.}'s~\cite{KapralovKSV17} resolution of $1$-pass streaming complexity in the \ea model of 
approximating {\sc Max Cut} in graphs and Assadi {\em et al.}'s~\cite{AssadiKL16} resolution of the $1$-pass streaming complexity in the \ea model of approximating {\sc Set Cover}. Assadi {\em et al.}~\cite{AssadiKL16} also showed an interesting dichotomy of approximating the size of the optimal set cover and outputting an approximately optimal set cover.} 
Fafianie and Kratsch~\cite{FafianieK14} were the first to study parameterized streaming complexity of NP-hard problems like $d$-{\sc Hitting Set} and {\sc Edge Dominating Set} in graphs. Chitnis {\em et al.}~\cite{ChitnisCEHMMV16,ChitnisCEHM15,ChitnisCHM15} developed a 
sampling technique to design efficient parameterized streaming algorithms for promised variants of {\sc Vertex Cover}, $d$-{\sc Hitting Set} problem, $b$-{\sc Matching} etc. They also proved lower bounds for problems like $\mathcal{G}$-{\sc Free Deletion}, $\mathcal{G}$-{\sc Editing}, {\sc Cluster Vertex Deletion} etc.~\cite{ChitnisCEHMMV16}.  

\smallskip 
\noindent
{\bf Organisation of the paper.}
{Section~\ref{sec:prelim-main} contains preliminary definitions. Our algorithm for \cvd is described in Section~\ref{sec:cvd}. 
The algorithms for \cn, \subdel and \minordel are given in Section~\ref{sec:minor}. The lower bound results are in Section~\ref{sec:lbproof}. Appendix~\ref{sec:probdefi} has all formal problem definitions.\remove{ The pseudocodes for the algorithms in Sections~\ref{sec:cvd} and~\ref{sec:minor} are presented in Appendix~\ref{sec:pseudocode}. Appendices~\ref{app:proofcvd} and~\ref{app:proofminor} contain the missing proofs from Sections~\ref{sec:cvd} and~\ref{sec:minor}, respectively. }}

\section{Preliminaries, Model and Relationship Between Models}\label{sec:prelim-main}
In this section we state formally the models of streaming algorithms we use in this paper, relationship between them and some preliminary notations that we make use of. 
\vspace{-0.4cm}
\paragraph*{Streaming Models.}
A promising prospect to deal with problems on large graphs is the study of \emph{streaming algorithms}, where a compact sketch of the subgraph whose edges have been \emph{streamed}/revealed so far, 
is stored and computations are done on this sketch. Algorithms that can access the sequence of edges of the input graph, $p$ times in the same order, are defined as \emph{$p$-pass streaming algorithms}. For simplicity, we refer to 1-pass streaming algorithms as \emph{streaming algorithms}. The space used by a ($p$-pass) streaming algorithm, is defined as the \emph{streaming complexity} of the algorithm.\remove{ Usually, the graph is accessed only once. For \emph{streaming algorithms}, it is the space complexity of the algorithm, or the \emph{streaming complexity}, that is optimized.} The algorithmic model to deal with streaming graphs is determined by the way the graph is revealed. 
Streaming algorithms for graph problems are usually studied in the following 
models~\cite{CormodeDK18,McGregor14,McGregorVV16}. For the upcoming discussion, $V(G)$ and $E(G)$ will denote the vertex and edge set, respectively of the graph $G$ having $n$ vertices.
\begin{itemize}[noitemsep, wide=0pt, leftmargin=\dimexpr\labelwidth + 2\labelsep\relax]
\item[(i)] {\eam(\ea) model:} The stream consists of edges of $G$ in an arbitrary order.
\item[(ii)] {\deam(\dea) model:} Each element of the input stream is a pair $(e,\mbox{state})$, where $e \in E(G)$ and $\mbox{state} \in \{\mbox{insert, delete}\}$  describes whether $e$ is being inserted into or deleted from the current graph. 
\item[(iii)] {\vam(\va) model:} The vertices of $V(G)$ are exposed in an arbitrary order. After a vertex $v$ is exposed, all the edges between $v$ and  neighbors of $v$ that have already been exposed, are revealed. This set of edges are revealed one by one in an arbitrary order.
\item[(iv)] {\alm(\al) model:} The vertices of $V(G)$ are exposed in an arbitrary order. When a vertex $v$ is exposed, all the edges that are incident to $v$, are revealed one by one in an arbitrary order. Note that in this model each edge is exposed twice, once for each exposure of an endpoint.
\end{itemize}
\vspace{-0.4cm}
\paragraph*{Streamability and Hardness.}
\noindent Let $\Pi$ be a parameterized graph problem that takes as input a graph on $n$ vertices and a parameter $k$. Let $f:\mathbb{N}\times \mathbb{N} \rightarrow \mathbb{R}$ be a \emph{computable} function. For a model $\cM \in \{\mbox{\dea, \ea, \va, \al}\}$, whenever we say that \emph{an algorithm $\cA$ solves $\Pi$ with complexity $f(n,k)$ in model $\cM$}, we mean $\cA$ is a randomized algorithm that for any input instance of $\Pi$ in model  $\cM$ gives the correct output with probability $2/3$ and has streaming complexity $f(n,k)$.   

\begin{defi}
\label{defi:hard}
A parameterized graph problem $\Pi$, that takes an $n$-vertex graph and a parameter $k$ as input, is \emph{$\Omega(f)$ $p$-pass hard} in the \eam model, or in short $\Pi$ is \emph{$(\ea,f,p)$-hard}, if there does not exist any $p$-pass streaming algorithm of streaming complexity $\Oh(f(n,k))$ bits that can solve $\Pi$ in model $\cM$. 
\end{defi}
Analogously, \emph{$(\dea,f,p)$-hard}, \emph{$(\va,f,p)$-hard} and \emph{$(\al,f,p)$-hard} are defined.

\begin{defi}
\label{defi:streamable}
A graph problem $\Pi$, that takes an $n$-vertex graph and a parameter $k$ as input, is \emph{$\Oh(f)$ $p$-pass streamable} in \eam model, or in short $\Pi$ is \emph{$(\ea,f,p)$-streamable} if there exists a $p$-pass streaming algorithm of streaming complexity $\Oh(f(n,k))$ words~\footnote{It is usual in streaming that the lower bound results are in bits, and the upper bound results are in words.} that can solve $\Pi$ in \eam model. 
\end{defi}
\emph{$(\dea,f,p)$-streamable}, \emph{$(\va,f,p)$-streamable} and $(\al,f,p)$-\emph{streamable} are defined analogously. For simplicity, we refer to $(\cM,f,1)$-hard and $(\cM,f,1)$-streamable as \emph{$(\cM,f)$}-hard and \emph{$(\cM,f)$-streamable}, respectively, where $\cM \in\{\dea, \ea, \va, \al\}$.

\remove{\todo[inline]{How do we prove this observation. It is better not to state such observations. To prove this we would need to formalize what problems are. All problems and all functions are vague terms. It is better to avoid such constructs. May be just say this all our problems and all computable function $f$ the following observation holds. May be a short proof in appendix will be helpful.}}

\begin{defi}
\label{defi:conn}
Let $\cM_1,\cM_2\in $ $\{\mbox{\dea,\ea,\va,\al}\}$ be two streaming models, $f:\N \times \N \rightarrow \R$ be a computable function, and $p \in \N$.  
\begin{itemize}[noitemsep, wide=0pt, leftmargin=\dimexpr\labelwidth + 2\labelsep\relax]
\item[(i)] If for any parameterized graph problem $\Pi $, $(\cM_1,f,p)$-hardness of $\Pi$ implies $(\cM_2,f,p)$-hardness of $\Pi$, then we say $\cM_1 \leq_h \cM_2$.
\item[(ii)]  If for any parameterized graph problem $\Pi $, $(\cM_1,f,p)$-streamability of $\Pi$ implies $(\cM_2,f,p)$-streamability of $\Pi$, then we say $\cM_1 \leq_s \cM_2$.
\end{itemize} 
\end{defi}

Now, from  Definitions~\ref{defi:hard}, \ref{defi:streamable} and \ref{defi:conn}, we have the following Observation.
\begin{obs}
\label{obs:vertex_edge_conn}
 $\mbox{\al}  \leq_h \mbox{\ea} \leq_h \mbox{\dea}$;
 $  \mbox{\va} \leq_h \mbox{\ea} \leq_h \mbox{\dea}$;
$\mbox{\dea} \leq_s \mbox{\ea} \leq_s \mbox{\va} $;
 $\mbox{\dea} \leq_s \mbox{\ea} \leq_s \mbox{\al}$.
\end{obs}
\noindent
This observation has the following implication. If we prove a lower (upper) bound result for some problem $\Pi$ in model $\cM$, then it also holds in any model $\cM'$ such that $\cM \leq_h \cM'$ ($\cM \leq_s \cM'$). For example, if we prove a lower bound result in \al or \va model, it also holds in \ea and \dea model; if we prove an upper bound result in \dea model, it also holds in \ea,~\va and \al model.
In general, there is no direct connection between \al and \va. In \al and \va, the vertices are exposed in an arbitrary order. However, we can say the following when the vertices arrive in a fixed (known) order.
\begin{obs}
\label{obs:vertex_edge_conn1}
 Let $\mbox{\al}'$ ($\mbox{\va}'$) be the restricted version of $\mbox{\al}$ ($\mbox{\va}$), where the vertices are exposed in a fixed (known) order. Then $\mbox{\al}'  \leq_h \mbox{\va}'$ and $\mbox{\va}' \leq_s \mbox{\al}'$. 
\end{obs}
Now, we remark the implication of the relation between different models discussed in this section to our results mentioned in Table~\ref{table:lb}. 
\begin{rem}
\label{rem:table}
In Table~\ref{table:lb}, the lower bound results in \va and \al hold even if we know the sequence in which vertices are exposed, and the upper bound results hold even if the vertices arrive in an arbitrary order. In general, the lower bound in the \al model for some problem $\Pi$ does not imply the lower bound in the \va model for $\Pi$. However, our lower bound proofs in the \al model hold even if we know the order in which vertices are exposed. So, the lower bounds for \fvs, \ect, \Oct in the \al model imply the lower bound in the \va model. By Observations~\ref{obs:vertex_edge_conn} and~\ref{obs:vertex_edge_conn1}, we will be done by showing a subset of the algorithmic and lower bound results mentioned in the Table~\ref{table:lb}.  

\end{rem}
\vspace{-0.4cm}
\paragraph*{General Notation.}
The set $\{1,\ldots,n\}$ is denoted as $[n]$. Without loss of generality, we assume that the number of vertices in the graph is $n$, which is a power of $2$. Given an integer $i \in [n]$ and $r \in [\log_2 n]$, $\bit(i,r)$ denotes the $r
\mbox{-{th}}$ bit in the bit expansion of $i$. The union of two graphs $G_1$ and $G_2$ with $V(G_1)=V(G_2)$, is $G_1\cup G_2$, where $V(G_1\cup G_2)=V(G_1)=V(G_2)$ and $E(G_1 \cup G_2)=E(G_1)\cup E(G_2)$. For $X \subseteq V(G)$, $G \setminus X$ is the  subgraph of $G$ induced by $V(G) \setminus X$. The degree of a vertex $u \in V(G)$, is denoted by $\mbox{{deg}}_G(u)$. The maximum and average degrees of the vertices in $G$ are denoted as $\Delta(G)$ and $\Delta_{av}(G)$, respectively. For a family of graphs $\cF$, $\Delta(\cF)=\max \limits _{F \in \cF } \Delta(F)$. A graph $F$ is a \emph{subgraph} of a graph $G$ if $V(F)\subseteq V(G)$ and $E(F)\subseteq E(G)$ be the set of edges that can be formed only between vertices of $V(F)$. A graph $F$ is said to be a \emph{minor} of a graph $G$ if $F$ can be obtained from $G$ by deleting edges and vertices  and by contracting edges. The neighborhood of a vertex $v \in V(G)$ is denoted by $N_G(v)$. For $S \subseteq V(G)$, $N_G(S)$ denotes the set of vertices in $V(G) \setminus S$ that are \emph{neighbors of every vertex in $S$}. A vertex $v \in N_G(S)$ is said to be a common neighbor of $S$ in $G$. The size of any minimum vertex cover in $G$ is denoted by $\vc(G)$. A cycle on the sequence of vertices $v_1,\ldots,v_n$ is denoted as $\cC(v_1,\ldots,v_n)$. For a matching $M$ in $G$, the vertices in the matching are denoted by $V(M)$. $C_t$ denotes a cycle of length $t$. $P_t$ denotes a path having $t$ vertices. A graph $G$ is said to a \emph{cluster} graph if $G$ is a disjoint union of cliques, that is, no three vertices of $G$ can form an induced $P_3$.

\remove{\section{The Algorithmic Results}\label{sec:alg}
\noindent
In this Section, we describe the algorithmic results. First, for \cvd parameterized by vertex cover size $K$, we show $(\dea,K^2\log^4 n)$-streamability. By Observation~\ref{obs:vertex_edge_conn}, this implies $(\cM,K^2\log^4 n)$-streamability for all $\cM \in \{\mbox{\ea,\va,\al}\}$. 
 Next, we that \subdel is $(\al,d \cdot K^{d+1})$-streamable, where $d=\Delta(\cF) \leq K$. As a corollary,  we show that \fvs, \ect, \Oct and \tdel parameterized by vertex cover size $K$  have $(\al,K^3)$-streamability. This complements the results in (a) and (b) of Theorem~\ref{theo:lowerbounds} (in Section~\ref{sec:lbproof}) that show that the problems parameterized by vertex cover size $K$ are $(\va,n/p,p)$-hard~(see also Table~1). Note that by Observation~\ref{obs:vertex_edge_conn}, this also implies that the problems parameterized by vertex cover size $K$ are $(\cM,n/p,p)$-hard when $\cM \in \{\mbox{\ea,\dea}\}$.
 }

\section{\cvd in the \dea model}
\label{sec:cvd}
\noindent
 In this Section, we show that \cvd parameterized by vertex cover size $K$, is $(\dea,K^2\log^4 n)$-streamable. By Observation~\ref{obs:vertex_edge_conn}, this implies $(\cM,K^2\log^4 n)$-streamability for all $\cM \in \{\mbox{\ea,\va,\al}\}$. The sketch of the algorithm for \cvd parameterized by vertex cover size $K$ in the \dea model is in Algorithm~\ref{algo_cvd}. The algorithm is inspired by the streaming algorithm for {\sc Vertex Cover}~\cite{ChitnisCEHMMV16}. Before discussing the algorithm, let us discuss some terms.

 A family of hash functions of the form $h:[n]\rightarrow [m]$ is said to be \emph{pairwise independent hash family} if for a pair $i,j \in [n]$ and a randomly chosen $h$ from the family, $\pr(h(i)=h(j))\leq \frac{1}{m}$. Such a hash function $h$ can be stored efficiently by using $\Oh(\log n )$ bits~\cite{MotwaniR95}. 

\noindent {\bf $\ell_0$-sampler}~\cite{CormodeF14}: Given a dynamic graph stream, an $\ell_0$-sampler does the following: with probability at least
$1-\frac{1}{n^c}$, where $c$ is a positive constant, it produces an edge uniformly at random from the set of edges that have been inserted so far but not deleted. If no such edge exists, $\ell_0$-sampler reports {\sc Null}.\remove{ This means that the $\ell_0$-sampler fails with probability at most $\frac{1}{n^c}$ to return an edge that has been inserted but not deleted.} The total space used by the sampler is $\Oh(\log ^3 n)$.

\begin{algorithm}[H]
\caption{\cvd}
\label{algo_cvd}
\KwIn{A graph $G$ having $n$ vertices in the \dea model,  with  vertex cover size at most $K \in \N$, solution parameter $k \in \N$, such that $k \leq K$.}
\KwOut{A set $X \subset V(G)$ of $k$ vertices such that $G \setminus X$ is a cluster graph if such a set exists. Otherwise, the output is {\sc Null}}
\Begin
	{
	
	 From a pairwise independent family of hash functions that map $ V(G)$ to $[\beta K]$, choose $h_1,\ldots,h_{\alpha \log n}$ such that each $h_i$ is chosen uniformly and independently at random, where $\alpha$ and $\beta$ are suitable \emph{large} constants. 
	
	For each $i \in [\alpha \log n]$ and $r,s \in [\beta K]$, initiate an $\ell_0$ sampler $L^i_{r,s}$.
	
	\For{ (each $(u,v)$ in the stream)}
	{
	Irrespective of $(u,v)$ being inserted or deleted, give the respective input to the $\ell_0$-samplers $L_{h_i(u),h_i(v)}^i$ for each $i \in [\alpha \log n]$.
	}
	
	For each $i \in [\alpha \log n]$, construct a 
subgraph $H_i$  by taking the outputs of all the $\ell_0$-samplers corresponding to the hash function $h_i$.

Construct $H =H_1 \cup \cdots \cup H_{\alpha \log n}$.

Run the classical {\sc FPT} algorithm for \cvd on the subgraph $H$ and solution size bound $k$~\cite{saketbook15}.

\If{($H$ has a solution $S$ of size at most $k$)}
{Report $S$ as the solution to $G$.}
\Else{
Report {\sc Null}}

}
\end{algorithm}

\begin{theo}
\label{theo:cvd_ub}
\cvd, parameterized by vertex cover size $K$, is $(\dea,K^2\log^4 n)$-streamable.
\end{theo}
\begin{proof}

Let $G$ be the input graph of the streaming algorithm and by assumption $\vc(G) \leq K$. 
Let $h_1,\ldots,h_{\alpha \log n}$ be a set of $\alpha \log n$ pairwise independent hash functions such that each $h_i$ chosen uniformly and independently at random from a pairwise independent family of hash functions, where $h:V(G) \rightarrow [\beta K]$, $\alpha$ and $\beta$ are suitable constants. For each hash function $h_i$ and pair $r,s \in [\beta K]$, let $G_{r,s}^i$ be the subgraph of $G$ induced by the vertex set $\{v \in V(G) : h_i(v) \in \{r,s\}\}$. For the hash function $h_i$ and for each pair $r,s \in [\beta K]$, we initiate an $\ell_0$ sampler for the dynamic stream restricted to the subgraph $G_{r,s}^i$. Therefore, there is a set of $\Oh(K^2)$ $\ell_0$-samplers $\{L^i_{r,s}: r,s \in [\beta K]\}$ corresponding to the hash function $h_i$. Now, we describe what our algorithm does when an edge is either 
inserted or deleted. A pseudocode of our algorithm for \cvd is given in Algorithm~\ref{algo_cvd}.
When an edge $(u,v)$ arrives in the stream, that is $(u,v)$ is inserted or deleted, we give the respective input to $L_{h_i(u),h_i(v)}^i$, where $i \in [\alpha \log n]$. At the end of the stream, for each $i \in [\alpha \log n]$, we construct a 
subgraph $H_i$ by taking the outputs of all the $\ell_0$-samplers corresponding to the hash function $h_i$.
Let $H =H_1 \cup \cdots \cup H_{\alpha \log n}$. We run the classical {\sc FPT} algorithm for \cvd on the subgraph $H$ and solution size bound $k$~\cite{saketbook15}, and report \yes to \cvd if and only if we get \yes as answer from the above {\sc FPT} algorithm on $H$. If we output \yes, then we also give the solution on $H$ as our solution to $G$. 

 The correctness of the algorithm needs an existential structural result on $G$ (Claim~\ref{clm:cvd_inter}) and the fact that if there exists a set $X \subset V(G)$ whose deletion turns $H$ into a cluster graph, then the same $X$ deleted from $G$ will turn it into a cluster graph with high probability (Claim~\ref{lem:cvd_ub_main}). 
 
\begin{cl}
\label{clm:cvd_inter}
There exists a partition $\mathcal{P}$ of $V(G)$ into $Z_1,\ldots,Z_t,I$ such that
 the subgraph induced in $G$ by each $Z_i$, is a clique with at least $2$ vertices, and the subgraph induced by $I$ is the empty graph.\remove{ Moreover, if $\vc(G) \leq K$, then $\size{C_i}\leq K+1$ , $t \leq K$ and $\size{\bigcup\limits_{i \in [t]}C_i} \leq 2K$.}
\end{cl}
\begin{proof}[Proof of Claim~\ref{clm:cvd_inter}]
We start with a partition which may not have the properties of the claim and modify it iteratively such that the final partition does have all the properties of the Claim. Let us start with a partition $\cP$ that does not satisfy the given condition. First, if there exists a part $Z_i$ having one vertex $v$, we create a new partition by adding $v$ to $I$. Next, if there exists a part $Z_i$ having at least two vertices and the subgraph induced by $Z_i$ is not a clique, then we partition $Z_i$ into smaller parts such that each smaller part is either a clique having at least two vertices or a singleton vertex\remove{possibly having some neighbors in $C_i$}. We create a new partition by replacing $Z_i$ with the smaller cliques of size at least $2$ and adding all the singleton vertices to $I$. Now, let $\cP'$ be the new partition of $V(G)$ obtained after all the above modifications. In $\cP'$, each part except $I$ is a clique of at least two vertices. If the subgraph induced by $I$ has no edges, $\cP'$ satisfies the properties in the Claim and we are done. Otherwise, there exists $u,v \in I$ such that $(u,v) \in E(G)$. In this case, we create a new part with $\{u,v\}$, and remove both $u$ and $v$ from $I$. Note that in the above iterative description, each vertex goes to a new part at most $2$ times - (i) it can move at most once from a part $Z_i$ to a smaller part $Z_j$ that is a clique on at least $2$ vertices and such a vertex will remain in the same part in all steps afterwards, or it can move at most once from a $Z_i$ to $I$, and (ii) a vertex can move at most once from $I$ to become a part of a clique $Z_i$ with at least $2$ vertices and such a vertex will remain in the same part in all steps after that. Therefore, this process is finite and there is a final partition that we obtain in the end. This final partition has all the properties of the claim.    
\end{proof}
\begin{cl}
\label{lem:cvd_ub_main}
Let $X \subset V(H)$ be such that $H \setminus X$ is a cluster graph. Then $G \setminus X$ is a cluster graph with high probability.
\end{cl}
\begin{proof}
Consider a partition $\cP$ of $V(G)$ into $Z_1,\ldots,Z_t,I$ as mentioned in Claim~\ref{clm:cvd_inter}. Note that our algorithm does not need to find such a partition. The existence of $\cP$ will be used only for the analysis purpose. Let $\cZ=\cup_{i=1}^t Z_i$. Note that since $\vc(G) \leq K$, each $Z_i$ can have at most $K+1$ vertices, and it must be true that $t\leq \vc(G) \leq K$. In fact, we can obtain the following stronger bound that $\vert \cZ \vert \leq 2K$. The total number of vertices in $\cZ$ is at most $\vc(G) + t$. Since $t \leq \vc(G) \leq K$, the total number of vertices in $\cZ$ is at most $2K$. 

 A vertex $u \in V(G)$, is said to be of \emph{high degree} if ${\deg}_G(u) \geq 40K$, and \emph{low degree}, otherwise. Let $V_h \subseteq V(G)$ be the set of all high degree vertices and $V_{\ell}$ be the set of low degree vertices in $G$. Let $E_{\ell}$ be the set of edges in $G$  having both the endpoints in $V_{\ell}$. It can be shown~\cite{ChitnisCEHMMV16} that
\begin{itemize}
\item[(i)] {\bf Fact-1:} $\size{V_h} \leq K$, $E_{\ell} = \Oh(K^2)$;
\item[(ii)] {\bf Fact-2:} $E_{\ell} \subseteq E(H)$, and ${\deg}_H(u) \geq 4K$ for each $u \in V_h$, with probability at least $1-\frac{1}{n^{\Oh(1)}}$.
\end{itemize}
Note that Fact-2 makes our algorithmic result for \cvd probabilistic.

Let $\cvdsol(G)  \subset V(G)$ denote a minimum set of vertices such that $G \setminus \cvdsol(G)$ 
is a cluster graph.  Our parametric assumption says that $\size{\cvdsol(G)} \leq \vc(G) \leq K$. Now consider the fact that a graph is a cluster graph if and only if it does not have any induced $P_3$. First, we show that the high degree vertices in $G$ surely need to be deleted to make it a cluster graph, i.e., $V_h \subseteq \cvdsol (G)$. Let us consider a vertex $u \in V_h$. 
 As the subgraph induced by $I$ has no edges and $\size{\cZ} \leq 2K$, each vertex in $I$ is of degree at most $\size{\cZ} \leq 2K$. So, $u$ must be in some $Z_i$ in the partition $\mathcal{P}$. As $\deg _G(u) \geq 40K$, using $\vert \cZ \vert  \leq 2K$, $u$ must have at least $38K$ many vertices from $I$ as its neighbors in $G$. Thus, there are at least $19K$ edge disjoint induced $P_3$'s that are formed with $u$ and its neighbors in $I$. If $u \notin {\cvdsol}(G)$, then more than $K$ neighbors of $u$ that are in $I$ must be present in $\cvdsol(G)$. It will contradict the fact that $\size{\cvdsol(G)} \leq \vc(G) \leq K$. Similarly, we can also argue that $V_h \subseteq \cvdsol(H)=X$ as $\deg _H(u) \geq 4K$ by Fact-2.

Next, we show that an induced $P_3$ is present in $G \setminus V_h$ if and only if it is present in $H \setminus V_h$. Removal of $V_h$ from $G~(\mbox{or}~H)$ removes all the induced  $P_3$'s in $G~(\mbox{or}~H)$ having at least one vertex in $V_h$. Any induced $P_3$ in $G \setminus V_h$ (or $H \setminus V_{h}$) must have all of its vertices as low degree vertices. Now, using Fact-2, note that all the edges, in $G$, between low degree vertices are in $H$. In other words, an induced $P_3$ is present in $G \setminus V_h$ if and only if it is present in $H \setminus V_h$. Thus for a set $X \subseteq V(G)$, if $(H \setminus V_h) \setminus X$ is a cluster graph then $(G\setminus V_h) \setminus X$ is also a cluster graph. 

Putting everything together, if $X \subseteq V(G)$ is such that $H \setminus X$ is a cluster graph, then $G \setminus X$ is also a cluster graph.
\end{proof}

Coming back to the proof of Theorem~\ref{theo:cvd_ub}, we are using $\Oh(\log n)$ hash functions, and each hash function requires a storage of $\Oh(\log n)$ bits.
 There are $\Oh(K^2)$ $\ell_0$-samplers for each hash function and each $\ell_0$-sampler needs $\Oh(\log^3 n)$ bits of storage. Thus, the total space used by our algorithm is $\Oh(K^2\log^4 n)$.
\end{proof}

\section{Deterministic algorithms in the \al model}
\label{sec:minor}
In this Section, we show that \subdel is $(\al,\Delta(\cF) \cdot K^{\Delta (\cF)+1})$-streamable when the vertex cover of the input graph is parameterized by $K$. This will imply that \fvs, \ect, \Oct and \tdel parameterized by vertex cover size $K$, are $(\al,K^3)$-streamable. This complements the results in Theorems~\ref{theo:lowerbounds1} and~\ref{theo:lowerbounds2}  (in Section~\ref{sec:lbproof}) that show that the problems parameterized by vertex cover size $K$ are $(\va,n/p,p)$-hard~(see also Table~1). Note that by Observation~\ref{obs:vertex_edge_conn}, this also implies that the problems parameterized by vertex cover size $K$ are $(\cM,n/p,p)$-hard when $\cM \in \{\mbox{\ea,\dea}\}$.
Finally, we design an algorithm for \minordel that is inspired by the algorithm 
for \subdel.

 For the algorithm for \subdel, we define an auxiliary problem \cn and a streaming algorithm for it. This works as a subroutine for our algorithm for \subdel.
 \subsection{\cn problem}

For a graph $G$ and a parameter $\ell \in \N$, $H$ will be called \emph{a common neighbor subgraph} for $G$ if 
\begin{itemize}[noitemsep,wide=0pt, leftmargin=\dimexpr\labelwidth + 2\labelsep\relax]
\item[(i)] $V(H) \subseteq V(G)$ such that $H$ has no isolated vertex;
\item[(ii)] $E(H)$ contains the edges
\begin{itemize}[noitemsep,wide=0pt, leftmargin=\dimexpr\labelwidth + 2\labelsep\relax]
\item  of a maximal matching $M$ of $G$ along with the edges where both the endpoints are from $V(M)$, 
\item such that for each subset $S \subseteq V(M)$, $\size{S} \leq d$, $\size{N_H(S) \setminus V(M)} = \min \{\size{N_G(S) \setminus V(M)},\ell\}$, that is, $E(H)$ contains edges to at most $\ell$ common neighbors of $S$ in $N_G(S) \setminus V(M)$. 
\end{itemize}
\end{itemize}
In simple words, a common neighbor subgraph $H$ of $G$ contains the subgraph of $G$ induced by $V(M)$ as a subgraph of $H$ for some maximal matching $M$ in $G$. Also, for each subset $S$ of at most $d$ vertices in $V(M)$, $H$ contains edges to \emph{sufficient} common neighbors of $S$ in $G$. 
 The parameters $d \leq K$ and $\ell$ are referred to as the degree parameter and common neighbor parameter, respectively. 
 
 \remove{Figure~\ref{fig:cngraph} shows an illustration of common neighbor subgraph.
 
 \begin{figure}
  \centering
  \includegraphics[width=1 \linewidth]{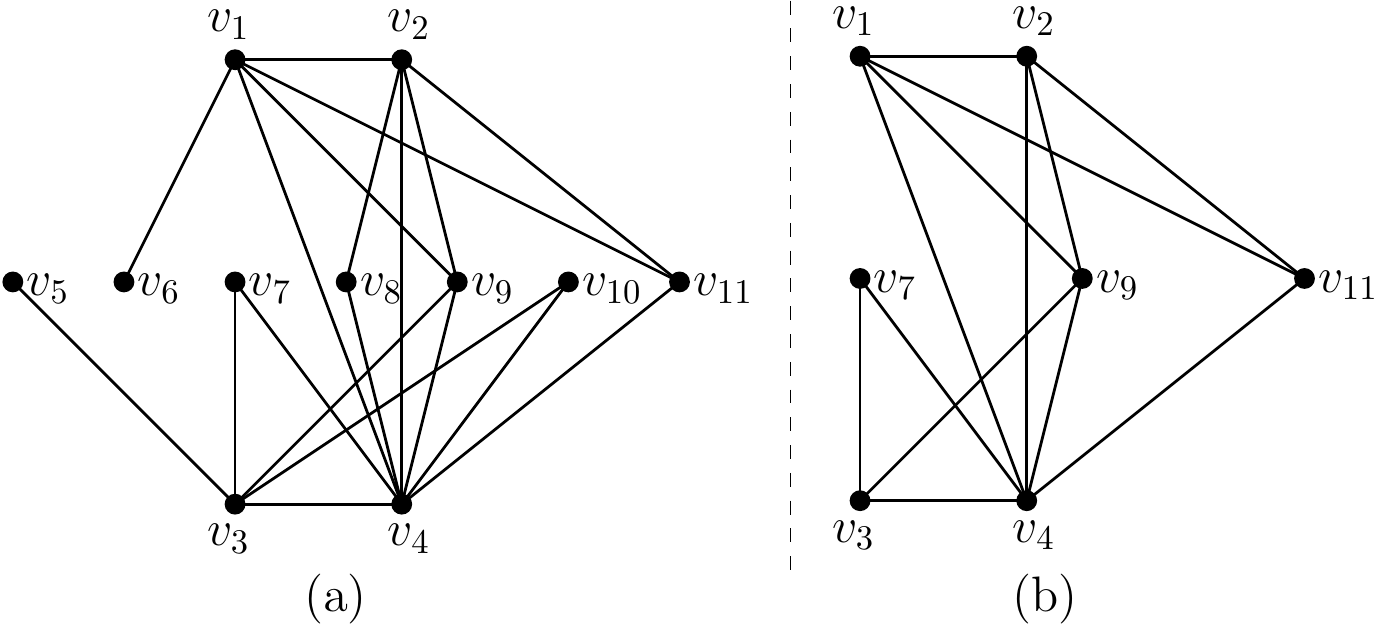}
  \caption{$(a)$ shows a graph $G$ where $M=\{(v_1,v_2),(v_3,v_4)\}$ is a maximal matching. $(b)$ shows the corresponding common neighbor subgraph $H$ with common neighbor parameter $\ell =2$.}
  \label{fig:cngraph}
\end{figure}}

The \cn problem is formally defined as follows. It takes as input a graph $G$ with $\vc(G) \leq K$, degree parameter $d \leq K$ and common neighbor parameter $\ell$ and produces a common neighbor subgraph of $G$ as the output. \cn parameterized by vertex cover size $K$, has the following result.

  \begin{algorithm}[H]
\caption{\cn}
\label{algo:cn1}
\KwIn{A graph $G$, with $\vc(G) \leq K$, in the \al model, a degree parameter $d \leq K$, and a common neighbor parameter $\ell$.}
\KwOut{A common neighbor subgraph $H$ of $G$.}
\Begin
	{
	Initialize $M = \emptyset$ and $V(M)=\emptyset$, where $M$ denotes the current maximal matching.
	
	Initialize a temporary storage $T =\emptyset$.
	
	\For{(each vertex $u \in V(G)$ exposed in the stream)}
	{

			\For{(each $(u,x) \in E(G)$ in the stream)}
			{
				\If{($u \notin V(M) $ and  $x \notin V(M)$)}
				{Add $(u,x)$ to $M$ and both $u,x$ to $V(M)$.}
				 
				 \If{($x \in V(M)$)}
				{ Add $(u,x)$ to $T$.}
			}
			\If{( If $u$ is added to $V(M)$ during the exposure of $u$)}
			{Add all the edges present in $T$ to $E(H)$.}
			\Else
			{ 
				\For{(each $S \subseteq V(M)$ such that $\size{S} \leq d$ and $(u,z) \in T ~\forall z \in S $)}
				{
				\If{($N_H(S)$ is less than $\ell$ )}
				{Add  the edges $(u,z)$ $\forall z \in S$ to $E(H)$.}
				}
			} 
			Reset $T$ to $\emptyset$.
			
	}
	
}
\end{algorithm}
 \begin{lem}
\label{lem:cn1}
\cn, with a commmon neighbor parameter $\ell$ and parameterized by vertex cover size $K$, is $(\al,K^2\ell)$-streamable. 
\end{lem}
\begin{proof}
We start our algorithm by initializing $M =\emptyset$ and construct a matching in $G$ that is maximal under inclusion; See Algorithm~\ref{algo:cn1}.\remove{extend the current maximal matching $M$ of $G$ in the usual way. That is whenever we get an edge $(a,b)$ in the stream, we check whether both $a,b \notin V(M)$. If yes, we replace $M$ by $M \cup \{a,b\}$. Otherwise, we keep $M$ as it is.} As $\size{\vc(G)} \leq K$, $\size{M} \leq K$. Recall that we are considering the \al model here. Let $M_u$ and $M'_u$ be the maximal matchings just before and after the exposure of the vertex $u$ (including the processing of the edges adjacent to $u$), respectively. Note that, by construction these partial matchings $M_u$ and $M'_u$ are also maximal matchings in the subgraph exposed so far. The following Lemma will be useful for the proof.

\begin{cl}
Let $u \in N_G(S) \setminus V(M)$ for some $S \subseteq V(M)$. Then $S \subseteq V(M_u)$, that is, $u$ is exposed, after all the vertices in $S$ are declared as vertices of $V(M)$.  
\end{cl}
\begin{proof}
Observe that if there exists $x \in S$ such that $x \notin V(M_u)$, then after $u$ is exposed, there exists $y \in N_G(u)$ such that $(u,y)$ is present in $M_u'$. This implies $u \in V(M_u') \subseteq V(M)$, which is a contradiction to $u \in N_G(S) \setminus V(M)$.
\end{proof}
Now, we describe what our algorithm does when a vertex $u$ is exposed. A complete pseudocode of our algorithm for \cn is given in Algorithm~\ref{algo:cn1}. 
When a vertex $u$ is exposed in the stream, we try to extend the maximal matching $M_u$. Also, we store all the edges  of the form $(u,x)$ such that $x \in V(M_u)$, in a temporary memory $T$. As $\size{M_u} \leq K$, we are storing at most $2K$ many edges in $T$. Now, there are the following possibilities.
\begin{itemize}[noitemsep,wide=0pt, leftmargin=\dimexpr\labelwidth + 2\labelsep\relax]
\item If $u \in V(M_u')$, that is, either $u \in V(M_u)$ or the matching $M_u$ is extended by one of the edges stored in $T$, then we add all the edges stored in $T$ to $E(H)$.
\item Otherwise, for each $S \subseteq V(M_u)$ such that $\size{S} \leq d$ and
 $S \subseteq N_G(u)$, we check whether the number of common neighbors of the vertices present in $S$, that are already stored, is less than $ \ell$. If yes, we add all the edges of the form $(u,z)$ such that $z \in S$ to $E(H)$; else, we do nothing. Now, we reset $T$ to $\emptyset$.
\end{itemize}

As $\size{M} \leq K$, $\size{V(M)} \leq 2K$. We are storing at most $\ell$ common neighbors for each $S \subseteq V(M)$ with $\size{S}\leq d$ and the number of edges having both the endpoints in $M$ is at most $\Oh(K^2)$, the total amount of space used is at most $\Oh(K^d\ell)$. 
\end{proof}

We call our algorithm described in the proof of Lemma~\ref{lem:cn1}
 and given in Algorithm~\ref{algo:cn1}, as $\cA_{cn}$. The following structural Lemma of the common neighbor subgraph of $G$, obtained by algorithm $\cA_{cn}$ is important for the design and analysis of streaming algorithms for \subdel. The proof of this structural result is similar to that in~\cite{FOMIN2014468}.

\begin{lem}
\label{lem:structure1}
 Let $G$ be a graph with $\vc(G) \leq K$ and let $F$ be a connected graph with $\Delta(F) \leq d \leq K$. Let $H$ be the common neighbor subgraph of $G$ with degree parameter $d$ and common neighbor parameter $(d+2) K$, obtained by running the algorithm $\cA_{cn}$. Then the following holds in $H$: For any subset $X \subseteq V(H)$, where $\vert X \vert \leq K$, $F$ is a subgraph of $G\setminus X$ if and only if $F'$ is a subgraph of $H\setminus X$, such that $F$ and $F'$ are \emph{isomorphic}.
\end{lem}
\begin{proof}
Let the common neighbor subgraph $H$, obtained by algorithm $\cA_{cn}$, contain a maximal matching $M$ of $G$. First, observe that since $\vc(G) \leq K$, the size of a subgraph $F$ in $G$ is at most $dK$. 
Now let us consider a subset $X \subseteq V(H)$ such that $\vert X \vert \leq K$. First, suppose that $F'$ is a subgraph of $H \setminus X$ and $F'$ is isomorphic to $F$. Then since $H$ is a subgraph of $G$, $F'$ is also a subgraph of $G \setminus X$. Therefore, $F = F'$ and we are done.

Conversely, suppose $F$ is a subgraph of $G \setminus X$ that is not a subgraph in $H \setminus X$. We show that there is a subgraph $F'$ of $H \setminus X$ such that $F'$ is isomorphic to $F$. Consider an arbitrary ordering $\{e_1,e_2,\ldots,e_s\} \subseteq (E(G) \setminus E(H)) \cap E(F)$; note that $s \leq \size{E(F)}$. We describe an iterative subroutine that converts the subgraph $F$ to $F'$ through $s$ steps, or equivalently, through a sequence of isomorphic subgraphs $F_0,F_1,F_2,\ldots F_s$ in $G$ such that $F_0=F$ and $F_s = F'$.

Let us discuss the consequence of such an iterative routine. Just before the starting of step $i \in [s]$, we have the subgraph $F_{i-1}$ such that $ F_{i-1 }$ is isomorphic to $F$ and the set of edges in $(E(G) \setminus E(H)) \cap E(F_{i-1})$ is a subset of $\{e_{i},e_{i+1},\ldots,e_s\}$. In step $i$, we convert the subgraph $F_{i-1}$ into $F_i$ such that $F_{i-1}$ is isomorphic to 
$F_i$. Just after the step $i \in [s]$, we have the subgraph $F_{i}$ such that $F_i$ is isomorphic to $F$ and the set of edges in $(E(G) \setminus E(H)) \cap E(F_i)$ is a subset of $\{e_{i+1},e_{i+2},\ldots,e_s\}$. In particular, in the end $F_s = F'$ is a subgraph both in $G$ and $H$.

Now consider the instance just before step $i$. We show how we select the subgraph $F_{i}$ from $F_{i-1}$. Let $e_{i}=(u,v)$. Note that $e_i \notin E(H)$.  By the definition of the maximal matching $M$ in $G$, it must be the case that $\vert \{u,v\} \cap V(M) \vert \geq 1$. From the construction of the common neighbor subgraph $H$, if both $u$ and $v$ are in $V(M)$, then $e_i=(u,v) \in E(H)$. So, exactly one of $u$ and $v$ is present in $V(M)$. Without loss of generality, let $u \in V(M)$. Observe that $v$ is a common neighbor of  $N_{G}(v)$ in $G$.  Because of the maximality of $M$, each vertex in $N_G(v)$ is present in $V(M)$. Now, as $(u,v) \notin E(H)$, $v$ is not a common neighbor of $N_G(v)$ in $H$. From the construction of the common neighbor subgraph, $H$ contains $(d+2) K$ common neighbors of all the vertices present in $N_G(v)$. Of these common neighbors, at most $(d+1)K$ common neighbors can be vertices in $X \cup F_i$. Thus, there is a vertex $v'$ that is a common neighbor of all the vertices present in $N_G(v)$ in $H$ such that $F_{i+1}$ is a subgraph that is isomorphic to $F_i$. Moreover, $(E(G) \setminus E(H)) \cap E(F_{i+1}) \subseteq \{e_{i+2},e_{i+3}\ldots,e_s\}$. Thus, this leads to the fact that there is a subgraph $F'$ in $H\setminus X$ that is isomorphic to the subgraph $F$ in $G \setminus X$.
 \end{proof}

\subsection{Streambality results for \subdel and \minordel}
Our result on \cn leads us to the following streamability result for \subdel and \minordel. We first discuss the result on \subdel, which is stated in the following theorem.

\begin{theo}
\label{theo:sub_ub}
\subdel parameterized by vertex cover size $K$ is $(\al,d \cdot K^{d+1})$-streamable, where $d=\Delta(\cF) \leq K$.
\end{theo}
\begin{proof}
 Let $(G,k,K)$ be an input for \subdel, where $G$ is the input graph, $k \leq K$ is the size of the solution of \subdel, and the parameter $ K$ is at least $\vc(G)$.
 
 Now, we describe the streaming algorithm for \subdel. First, we run the \cn streaming algorithm described in Lemma~\ref{lem:cn1} (and given in Algorithm~\ref{algo:cn1}) with degree parameter $d$ and common neighbor parameter $(d +2) K$, and let the common neighbor subgraph obtained be $H$. We run a traditional FPT algorithm for \subdel~\cite{saketbook15} on $H$ and output \yes if and only if the output on $H$ is YES.

 Let us argue the correctness of this algorithm. By Lemma~\ref{lem:structure1}, for any subset $X \subseteq V(H)$, where $\vert X \vert \leq K$,  $F \in \cF$ is a subgraph of $G\setminus X$ if and only if  $F'$, such that $F'$ is isomorphic to $F'$, is a subgraph of $H\setminus X$. In particular, let $X$ be a $k$-sized vertex set of $G$. As mentioned before, $k \leq K$.
 Thus, by Lemma~\ref{lem:structure1}, $X$ is a solution of \subdel in $H$ if and only if $X$ is a solution of \subdel in $G$. Therefore, we are done with the correctness of the streaming algorithm for \subdel.
 
 The streaming complexity of \subdel is same as the streaming complexity for the algorithm $\cA_{cn}$ from Lemma~\ref{lem:cn1} with degree parameter $d=\Delta(\cF)$ and common neighbor parameter $(d+2)K$. Therefore, the streaming complexity of \subdel is $\Oh(d \cdot K^{d+1})$.
\end{proof}

\begin{coro}
\label{theo:cycle_ub}
\fvs, \ect, \Oct and \tdel parameterized by vertex cover size $K$ are $(\al,K^3)$-streamable due to deterministic algorithms.
\end{coro}
\subsection{Algorithm for \minordel}

Finally, we describe a streaming algorithm for \minordel that works similar to that of \subdel due to the following proposition and the result is stated in Theorem~\ref{theo:minor_ub}.

\begin{pre}[\cite{FOMIN2014468}]
\label{prop:minor-to-sub}
Let $G$ be a graph with $F$ as a minor and $\vc(G) \leq K$. Then there exists a subgraph $G^*$ of 
$G$ that has $F$ as a minor such that $\Delta(G^*) \leq \Delta(F)$ and $V(G^*) \leq V(F)+K(\Delta(F)+1) .$ 
\end{pre}
\begin{theo}
\label{theo:minor_ub}
\minordel parameterized by vertex cover size $K$ are $(\al,d \cdot K^{d+1})$-streamable, where $d=\Delta(\cF) \leq K$. 
\end{theo}
\begin{proof}
 Let $(G,k,K)$ be an input for \minordel, where $G$ is the input graph, $k$ is the size of the solution of \minordel we are looking for, and the parameter $K$ is such that $\vc(G) \leq K$. Note that, $k \leq K$.
 
 Now, we describe the streaming algorithm for \minordel. First, we run the \cn streaming algorithm described in Lemma~\ref{lem:cn1} with degree parameter $d$ and common neighbor parameter $(d +2) K$, and let the common neighbor subgraph obtained be $H$. We run a traditional FPT algorithm for \minordel~\cite{saketbook15} and output \yes if and only if the output on $H$ is YES.

 Let us argue the correctness of this algorithm, that is, we prove the following for any $F \in \cF$. 
$G \setminus X$ contains $F$ as a minor if and only if $H \setminus X $ contains $F'$ as a minor such that $F$ and $F'$ are isomorphic, where $X \subseteq V(G)$ is of size at most $K$.  For the only if part, suppose $H \setminus X$ contains $F'$ as a minor. Then since $H$ is a subgraph of $G$, $G \setminus X $ contains $F'$ as a minor. For the 
 if part, let $G \setminus X$ contains $F$ as a minor. By Proposition~\ref{prop:minor-to-sub}, $G \setminus X$ conatins a subgraph $G^{*}$ such that $G^*$ contains $F$ as a minor and $\Delta(G^*) \leq \Delta(F)$. Now, Lemma~\ref{lem:structure1} implies that $H \setminus X$ also contains a subgraph $\hat{G^*}$ that is isomorphic to $G^*$. Hence, $H \setminus X$ contains 
 $F'$ as a monor such that $F'$ is isomorphic to $F$. 

 The streaming complexity of the streaming algorithm for \minordel is same as the streaming complexity for the algorithm $\cA_{cn}$ from Lemma~\ref{lem:cn1} with degree parameter $d=\Delta(\cF)$ and common neighbor parameter $(d+2)K$. Therefore, the streaming complexity for \minordel is $\Oh(d \cdot K^{d+1})$.
\end{proof}

\section{The Lower Bounds}
\label{sec:lbproof}
\noindent
Before we prove the lower bound results presented in Table~\ref{table:lb}, note that a lower bound on {\sc Feedback Vertex Set} is also a lower bound for \subdel (deletion of cycles as subgraphs) and \minordel (deletion of 3-cycles as minors). Thus, we will be done by proving the following theorems; Observations~\ref{obs:vertex_edge_conn} and~\ref{obs:vertex_edge_conn1} imply the other hardness results.
\remove{\begin{theo}
\label{theo:lowerbounds}
\begin{itemize}
\item[(A)]
{\sc Feedback Vertex Set}, {\sc Even Cycle Transversal} and {\sc Odd Cycle Transversal} are (I) $(\al, n \log n)$-hard parameterized by solution size $k$ and even if $\Delta_{av}(G)=\Oh(1)$, (II) $(\al, n/p,p)$-hard parameterized by solution size $k$ and even if $\Delta(G)=\Oh(1)$, and (III) $(\va, n/p,p)$-hard parameterized by vertex cover size $K$ and even if $\Delta_{av}(G)=\Oh(1)$ . 

\item[(B)]
 \tdel is (I) $(\va, n \log n)$-hard parameterized by solution size $k$ and even if $\Delta_{av}(G)=\Oh(1)$, (II) $(\va, n/p,p)$-hard parameterized by solution size $k$ and even if $\Delta(G)=\Oh(1)$,  and (III) $(\va, n/p,p)$-hard parameterized by vertex cover size $K$ and even if $\Delta_{av}(G)=\Oh(1)$.

\item[(C)]
\cvd is  $(\va, n/p,p)$-hard parameterized by solution size $k$ and even if $\Delta(G)=\Oh(1)$. 
\end{itemize}
\end{theo}
}
\begin{theo}
\label{theo:lowerbounds1}
{\sc Feedback Vertex Set}, {\sc Even Cycle Transversal} and {\sc Odd Cycle Transversal} are 
\begin{itemize}
\item[(I)] $(\al, n \log n)$-hard parameterized by solution size $k$ and even if $\Delta_{av}(G)=\Oh(1)$, \item[(II)] $(\al, n/p,p)$-hard parameterized by solution size $k$ and even if $\Delta(G)=\Oh(1)$, and \item[(III)] $(\va, n/p,p)$-hard parameterized by vertex cover size $K$ and even if $\Delta_{av}(G)=\Oh(1)$ . 
\end{itemize}
\end{theo}

\begin{theo}
\label{theo:lowerbounds2}
 \tdel is
\begin{itemize}
\item[(I)] $(\va, n \log n)$-hard parameterized by solution size $k$ and even if $\Delta_{av}(G)=\Oh(1)$, \item[(II)] $(\va, n/p,p)$-hard parameterized by solution size $k$ and even if $\Delta(G)=\Oh(1)$,  and
\item[(III)] $(\va, n/p,p)$-hard parameterized by vertex cover size $K$ and even if $\Delta_{av}(G)=\Oh(1)$.
\end{itemize}
\end{theo}

\begin{theo}
\label{theo:lowerbounds3}
\cvd is  $(\va, n/p,p)$-hard parameterized by solution size $k$ and even if $\Delta(G)=\Oh(1)$.
\end{theo}
We prove the above theorems by reduction from communication complexity problems discussed below.
\subsection{Communication complexity results}
\label{append-sec:prelim}

\remove{
\paragraph*{General Notation.}
The set $\{1,\ldots,n\}$ is denoted in short by $[n]$. In this paper, without loss of generality, we assume that the input graph size $n$ is a power of $2$ and the bit expansion of $n$ is $\log_2 n$ many consecutive zeros~\footnote{For example: Take $n=32$. The bit expansion of $32$ is $100000$. We ignore the bit $1$ and say that the bit expansion of $32$ is $00000$.}. 
 We consider undirected graphs and denote an undirected graph as $G$. The vertex set of $G$ is denoted as $V(G)$ while the edge set is denoted as $E(G)$. The neighborhood of a vertex $v \in V(G)$ is denoted by $N_G(u)$. The maximum degree of any vertex in $G$ is denoted as $\Delta(G)$.
The average degree of vertices in $G$ is denoted by $\Delta_{av}(G)$. For a pair $u,v \in V(G)$, $N_G(u,v)$ denotes the number of vertices that are adjacent in $G$ to both $u$ and $v$. The size of any minimum vertex cover in $G$ is denoted by $\vc(G)$. A cycle on the sequence of vertices $v_1,\ldots,v_n$ is denoted as $\cC(v_1,\ldots,v_n)$. For a matching $M$ in $G$, the vertices in the matching are denoted by $V(M)$.
 Given an integer $i$ and its binary representation, $\bit(i,p)$ denotes the $p^{\tiny{th}}$ bit of the binary representation of $i$.
}

Lower bounds of communication complexity have been used to provide lower bounds for the  streaming complexity of problems.
 In Yao's two party communication model, Alice and Bob get
inputs and the objective is to compute a function of their inputs with minimum bits of communication. In one way communication, only Alice is allowed to send messages and Bob produces the final output; whereas in two way communication both Alice and Bob can send messages. 
\begin{defi}
\label{def:comm}
The \emph{one (two) way communication complexity} of a problem $\Pi$ is the minimum number of bits that must be sent by Alice to Bob (exchanged between Alice and Bob) to  solve $\Pi$ on any arbitrary input with success probability $2/3$. 
\end{defi}
The following problems are very fundamental problems in communication complexity and we use these problems in showing lower bounds on the streaming complexity of problems considered in this paper. 
\begin{itemize}[noitemsep,wide=0pt, leftmargin=\dimexpr\labelwidth + 2\labelsep\relax]
\item[(i)]  \ind : Alice gets as input ${\bf x} \in \{0,1\}^n$ and Bob has an index $j \in [n]$. Bob  wants to determine whether ${\bf x}_j=1$. Formally, $\indxy =1$ if ${\bf x}_j=1$ and $0$, otherwise. 
\item[(ii)] \disj : Alice and Bob get inputs ${\bf x}, {\bf y} \in \{0,1\}^n$, respectively. The objective is to decide whether there exists an $i \in [n]$ such that $x_i=y_i=1$. Formally, $\disjxy =0$ if there exists an $i \in [n]$ such that $x_i=y_i=1$ and $1$, otherwise. 
\item[(iii)] \perm~\cite{SunW15} : Alice gets a permutation $\pi : [n]\rightarrow [n]$ and Bob gets an index $j \in [n \log n]$. The objective of Bob is to decide the value of  $\permxy$, defined as the $j$-{th} bit in the string of $0$'s and $1$'s obtained by concatenating the bit expansions of $\pi(1)\ldots \pi(n)$. In other words, let $\Phi:[n \log n] \rightarrow [n] \times [\log n]$ be a bijective function defined as \\
$\Phi(j)=\left( \lceil \frac{j}{\log n} \rceil, j + \log n -\lceil \frac{j}{\log n} \rceil \times \log n  \right)$. For a permutation $\pi: [n] \rightarrow [n]$, Bob needs to determine the value of the $\gamma$-th bit of $\pi \left( \lceil \frac{j}{\log n} \rceil \right)$, where $\gamma={ \left(j + \log n -\lceil \frac{j}{\log n} \rceil \times \log n  \right)}$.

\end{itemize}

\begin{pre}[\cite{KushilevitzN97,SunW15}]\label{pre:cc}
\begin{itemize}[noitemsep,wide=0pt, leftmargin=\dimexpr\labelwidth + 2\labelsep\relax]
\item[(i)]   The one way communication complexity of \ind is $\Omega(n)$.
\item[(ii)]  The two way communication complexity of \disj is $\Omega(n)$.
\item[(iii)] The one way communication complexity of \perm is $\Omega(n \log n)$.
\end{itemize}
\end{pre}

\paragraph{A note on reduction from \ind, \disj, \perm :}
 A reduction from a problem $\Pi_1$ in one/two way communication complexity to a problem $\Pi_2$ in streaming algorithms is typically as follows: The two players Alice and Bob device a communication protocol for $\Pi_1$ that uses a streaming algorithm for $\Pi_2$ as a subroutine. Typically in a round of communication, a player gives inputs to the input stream of the streaming algorithm, obtains the compact sketch produced by the streaming algorithm  and communicates this sketch to the other player. This implies that a lower bound on the communication complexity of $\Pi_1$ also gives a lower bound on the streaming complexity of $\Pi_2$.
 
 The following Proposition summarizes a few important consequences of reductions from problems in communication complexity to problems for streaming algorithms:

 \begin{pre}\label{pre:cc2sc}
 \begin{itemize}[noitemsep,wide=0pt, leftmargin=\dimexpr\labelwidth + 2\labelsep\relax]
\item[(i)] If we can show a reduction from \ind to a problem $\Pi$ in model $\cM$ such that the reduction uses a $1$-pass streaming algorithm of $\Pi$ as a subroutine, then $\Pi$ is $(\cM,n)$-hard.
\item[(ii)] If we can show a reduction from \disj to a problem $\Pi$ in model $\cM$  such that the reduction uses a $1$-pass streaming algorithm of $\Pi$ as a subroutine, then $\Pi$ is $(\cM,n/p,p)$-hard, for any $p \in \N$~\cite{ChitnisCEHM15,BishnuGMS18,AgarwalMPVZ06}.
\item[(iii)] If we can show a reduction from \perm to a problem $\Pi$ in model $\cM$ such that the reduction uses a $1$-pass streaming algorithm of $\Pi$ as a subroutine, then $\Pi$ is 
$(\cM,n \log n)$-hard.
\end{itemize}
\end{pre}

\subsection{Proofs of Theorems~\ref{theo:lowerbounds1},~\ref{theo:lowerbounds2},~\ref{theo:lowerbounds3}}
\begin{proof}[{\bf Proof of Theorem~\ref{theo:lowerbounds1}}]
The proofs for all three problems are similar. We first consider {\sc Feedback Vertex Set}. To begin with, we show the hardness results of \fvs for solution size $k=0$.\remove{ For all three hardness results, we give reductions from problems in communication complexity.}

\begin{figure}[h!]
  \centering
  \includegraphics[width=0.9 \linewidth]{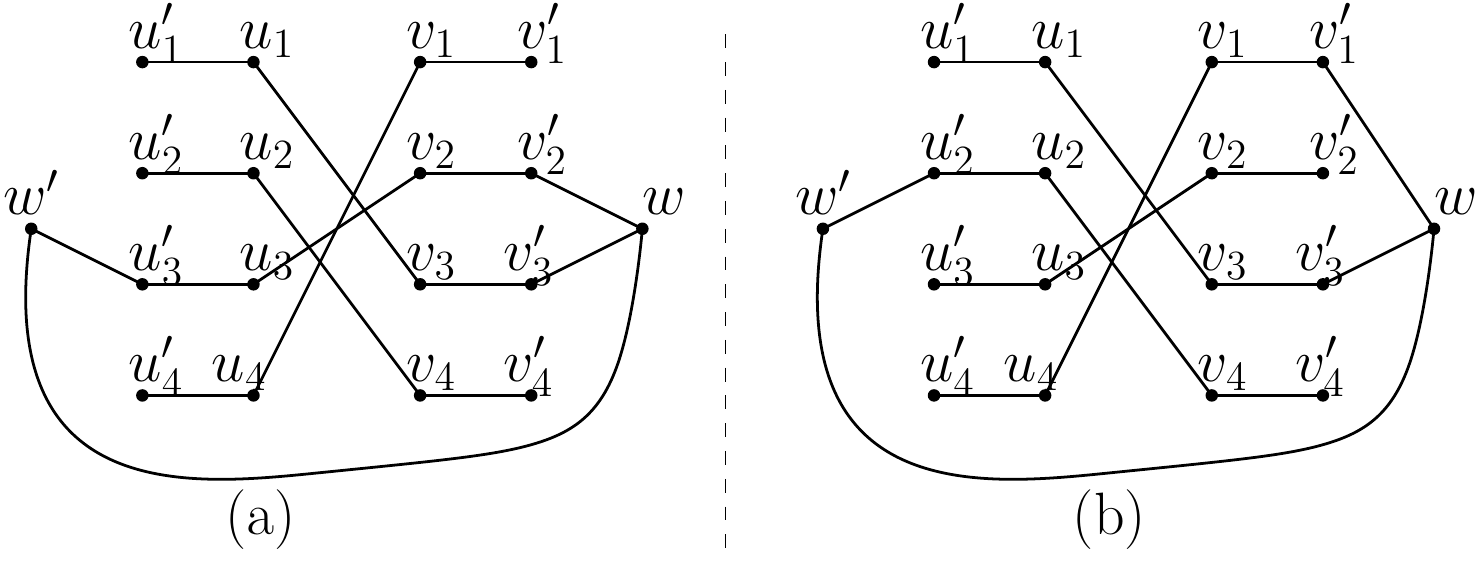}
  \caption{Illustration of Proof of Theorem~\ref{theo:lowerbounds1}~(I). Consider $n=4$. Let $\pi:[4]\rightarrow [4]$ such that $\pi(1)=3, \pi(2)=4, \pi(3)=2$ and $\pi(4)=1$. So the concatenated bit string is $11001001^{{2}}$. In (a), $j=5$, $\Phi(j)=(\psi,\gamma)=(3,1)$, $\permxy=1$, and $G$ contains a cycle. In (b), $j=4$, $\Phi(j)=(\psi,\gamma)=(2,2)$, $\permxy=0$, and $G$ does not contain a cycle.}
  \label{fig:fvs1}
\end{figure}
\footnotetext{Recall that we take $n$ as a power of $2$. For $1 \leq i \leq n-1$, the bit expansion of $i$ is the usual bit notation of $i$ using $\log_2 n$ bits; the bit expansion of $n$ is $\log_2 n$ many consecutive zeros. For example: Take $n=32$. The bit expansion of $32$ is $100000$. We ignore the bit $1$ and say that the bit expansion of $32$ is $00000$.}
\begin{proof}[Proof of Theorem~\ref{theo:lowerbounds1}~(I)] 
We give a reduction from \perm to \fvs in the \al model when the solution size parameter $k=0$.  The idea is to build a graph $G$ with $\Delta_{av}(G) =\Oh(1)$ and construct edges according to the input of \perm, such that the output of \perm is $0$ if and only if $G$ is cycle-free.

Let $\cA$ be a one pass streaming algorithm that solves \fvs in \al model using $o(n \log n)$ space. \remove{Now we give a protocol that solves \perm using $o(n \log n)$ space. }Let $G$ be a graph with $4n+2$ vertices $u_1,\ldots,u_n,$ $v_1,\ldots,v_n, u'_1, \dots,$ $u'_n, v'_1,\dots, v'_n, w,$ $w'$. Let $\pi$ be the input of Alice for \perm. See Figure~\ref{fig:fvs1} for an illustration.

{\bf Alice's input to $\cA$:} Alice inputs the graph $G$ first by exposing the vertices $u_1,\ldots,u_n, v_1,\ldots,$ $v_n$, sequentially. 
(i) While exposing the vertex $u_i$, Alice gives as input to $\cA$ the edges $(u_i,u'_i),(u_i, v_{\pi(i)})$; 
(ii) while exposing the vertex $v_i$, Alice gives the edges $(v_i,v'_i),(v_i,$ $ u_{\pi^{-1}(i)})$ to the input stream of $\cA$.

After the exposure of $u_1,\ldots,u_n, v_1,\ldots,$ $v_n$ as per the \al model, Alice sends the current memory state of $\cA$, i.e the sketch generated by $\cA$, to Bob. Let $j \in [n \log n]$ be the input of Bob and let $(\psi,\gamma) =\Phi(j)$.

{\bf Bob's input to $\cA$:}
 Bob exposes the vertices $u'_1 \dots, u'_n, v'_1,$ $\ldots,$ $ v'_n,w,w'$, sequentially. 
 (i) While exposing a vertex $u'_i$ where $i \neq \psi$, Bob gives the edge $(u'_i,u_i)$ to the input stream of $\cA$; 
 (ii) while exposing $u'_\psi$, Bob gives the edges $(u'_\psi,u_\psi)$ and $(u'_\psi,w')$;  
 (iii) while exposing a vertex $v'_i$, Bob gives the edge $(v'_i,v_i)$, and the edge $(v'_i,w)$ if and only if $\bit(i,\gamma)=1$;
 (iv) while exposing $w$, Bob gives the edge $(w,w')$, and the edge $(w,v'_i)$ if and only if $\bit(i,\gamma)=1$; 
 (v) while exposing $w'$, Bob gives the edges $(w',w)$ and $(w',u'_{\psi})$. 

 Observe that $\Delta_{av}(G)=\Oh(1)$. Now we show that the output of \fvs{} is \no if and only if $\permxy=1$. Recall that $k=0$.


From the construction, observe that $(w,w'), (w',u'_\psi), (u'_\psi,u_\psi),$ $(u_\psi, v_{\pi(\psi)}), (v_{\pi(\psi)}, v'_{\pi(\psi)}) \in E(G)$.
 When $\permxy =1$, the edge $(v'_{\pi(\psi)},w)$ is present in $G$. 
 So, $G$ contains the cycle $\cC (w,w',u'_{\psi},$ $u_\psi,v_{\pi(\psi)},v'_{\pi(\psi)} )$, that is, the output of \fvs is NO.
 
 On the other hand, if the output of \fvs is NO, then there is a cycle in $G$. From the construction, the cycle is 
 $\cC(w,w',u'_{\psi},u_\psi,v_{\pi(\psi)},v'_{\pi(\psi)})$. As $(v'_{\pi(\psi)},w)$ is an edge, the $\gamma$-{{th}} bit of $\pi(\psi)$ is $1$, that is $\permxy =1$.
 Now by Propositions~\ref{pre:cc} and \ref{pre:cc2sc}(iii), we obtain that {\sc Feedback Vertex Set} is $(\al, n \log n)$-hard even if $\Delta_{av}(G)=\Oh(1)$ and when $k=0$.
 \end{proof}
 \begin{figure}[h!]
  \centering
  \includegraphics[width=0.7\linewidth]{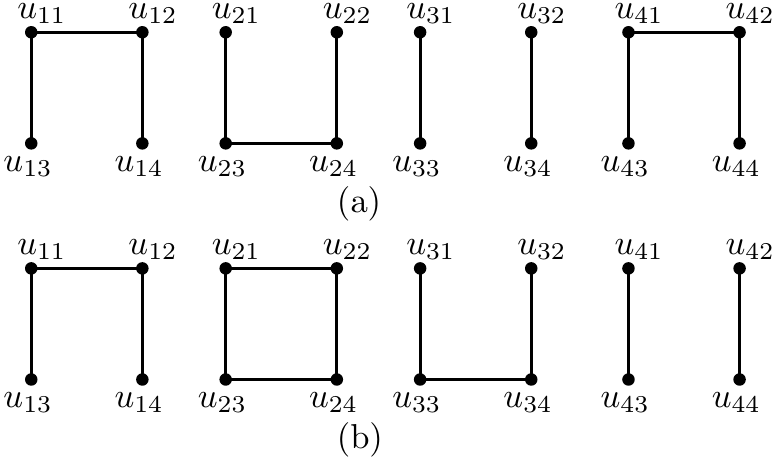}
  \caption{Illustration of Proof of Theorem~\ref{theo:lowerbounds1}~(II). Consider $n=4$. In (a), $\bx=1001$ and $\by=0100$, that is, $\disjxy=1$, and $G$ does not contain a cycle. In (b), $\bx=1100$ and $\by=0110$, that is, $\disjxy=0$, and $G$ contains a cycle.}
  \label{fig:fvs2}
\end{figure}   
 
\begin{proof}[Proof of Theorem~\ref{theo:lowerbounds1}~(II)] 
We give a reduction from \disj to \fvs in the \al model when the solution size parameter $k=0$.  The idea is to build a graph $G$ with $\Delta(G) =\Oh(1)$ and construct edges according to the input of \disj, such that the output of \disj is $1$ if and only if $G$ is cycle-free.

 Let $\cA$ be a one pass streaming algorithm that solves \fvs in \al model, such that $\Delta(G)=\Oh(1)$, and the space used is $o(n)$. Let $G$ be a graph with $4n$ vertices $u_{11},u_{12},u_{13},u_{14}, \ldots, u_{n1},$ $u_{n2},u_{n3},u_{n4}$. Let 
${\bf x,y}$ be the input of Alice and Bob for \disj, respectively. See Figure~\ref{fig:fvs2} for an illustration.

{\bf Alice's input to $\cA$:} Alice inputs the graph $G$ by exposing the vertices $u_{11},u_{12},u_{21},u_{22} \ldots, 
u_{n1},$ $u_{n2}$, sequentially. (i) While exposing $u_{i1}$, Alice gives as input to $\cA$ the edge $(u_{i1},u_{i3})$. Also, Alice gives the edge $(u_{i1},u_{i2})$ as input to $\cA$ if and only if ${x_i=1}$; (ii) while exposing $u_{i2}$, 
Alice gives the edge $(u_{i2},u_{i4})$ as input to $\cA$. Also, Alice gives the edge $(u_{i2},u_{i1})$ as input to $\cA$ if and only if $ x_i=1$. 

After the exposure of $u_{11},u_{12},u_{21},u_{22} \ldots, 
u_{n1},$ $u_{n2}$ as per the \al model, Alice sends current memory state of $\cA$, i.e. the sketch generated by $\cA$, to Bob. 

{\bf Bob's input to $\cA$:} Bob exposes the vertices $u_{13},u_{14},u_{23},$ $u_{24} \ldots,$ $ u_{n3},u_{n4}$ sequentially. (i) While exposing $u_{i3}$, Bob gives the edge $(u_{i3},u_{i1})$ as input to $\cA$, and gives the edge $(u_{i3},u_{i4})$ if and only if ${ y_i}=1$; (ii) while exposing $u_{i4}$, Bob gives the edge $(u_{i4},u_{i2})$ as input to $\cA$, and gives the edge $(u_{i4},u_{i3})$ if and only if $ y_i=1$. 

 Observe that $\Delta(G)\leq 4$. Recall that $k=0$. Now we show that the output of \fvs{} is \no if and only if $\disjxy =0$.

From the construction, $(u_{i1},u_{i3}), (u_{i2},u_{i4}) \in E(G)$, for each $i \in [n]$. If $\disjxy=0$,
 there exists $i \in [n]$ such that $x_i=y_i=1$. This implies the edges $(u_{i1},u_{i2})$  and $(u_{i3},u_{i4})$ are present in $G$. So, the cycle $\cC(u_{i1},u_{i2},u_{i3},u_{i4})$ is present in $G$, that is, the output of \fvs is NO.
 
Conversely, if the output of \fvs is NO, there exists a cycle in $G$. From the construction, the cycle must be 
 $\cC(u_{i1},u_{i2},$ $u_{i3},u_{i4})$ for some $i \in [n]$. As the edges $(u_{i1},u_{i2})$ and $(u_{i3},u_{i4})$ are present in $G$, $x_i=y_i=1$, that is, $\disjxy=0$. 

  Now by Propositions~\ref{pre:cc} and \ref{pre:cc2sc}(ii), we obtain that {\sc Feedback Vertex Set} is $(\al, n/p,p)$-hard even if $\Delta(G)=\Oh(1)$ and when $k=0$.
 \end{proof}

\begin{proof}[Proof of Theorem~\ref{theo:lowerbounds1}~(III)]
We give a reduction from \disj to \fvs in the \va model when the solution size parameter $k=0$.  The idea is to build a graph $G$ with vertex cover size bounded by $K$ and $\Delta(G)=\Oh(1)$, and construct edges according to the input of \disj, such that the output of \disj is $1$ if and only if $G$ is cycle-free.

 \begin{figure}[h!]
  \centering
  \includegraphics[width=0.8\linewidth]{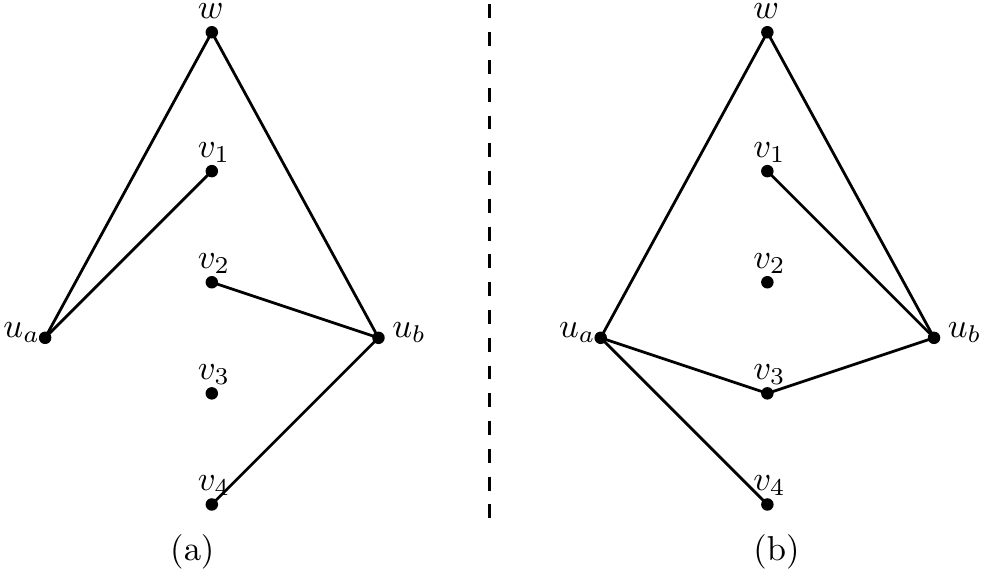}
  \caption{Illustration of Proof of Theorem~\ref{theo:lowerbounds1}~(III). Consider $n=4$. In (a), $\bx=1000$ and $\by=0101$, that is, $\disjxy=1$, and $G$ does not contain a cycle. In (b), $\bx=0011$ and $\by=1010$, that is, $\disjxy=0$, and $G$ contains a cycle. }
  \label{fig:fvs3}
\end{figure}

 Let $\cA$ be a one pass streaming algorithm that solves \fvs in \va model, such that $\vc(G) \leq K$ and $\Delta_{av}(G) =\Oh(1)$, and the space used is $o(n)$. Let $G$ be a graph with $n+3$ vertices $u_a,v_1,\ldots,v_n,u_b,w$. Let 
${\bf x,y}$ be the input of Alice and Bob for \disj, respectively. See Figure~\ref{fig:fvs3} for an illustration.

{\bf Alice's input to $\cA$:} Alice inputs the graph $G$ first by exposing the vertices $u_a, v_1,\ldots,v_n$, sequentially. (i) While exposing $u_a$, Alice does not give any edge; (ii) while exposing $v_i$, Alice gives the  
edge $(v_i,u_a)$, as input to $\cA$, if and only if $x_i=1$. 

After the exposure of $u_a, v_1,\ldots,v_n$ as per \va model, Alice sends the current memory state of $\cA$, i.e., the sketch generated by $\cA$, to Bob.

{\bf Bob's input to $\cA$:} Bob first exposes $u_b$ and then exposes $w$. (i) While exposing $u_b$, Bob gives the
edge $(u_b,v_i)$ if and only if $y_i=1$; (ii) while exposing $w$, Bob gives the edges $(w,u_a)$
and $(w,u_b)$, as inputs to $\cA$.

 From the construction, observe that $\vc(G) \leq 2 \leq K$ and $\Delta_{av}(G) = \Oh(1)$. Recall that $k=0$.
Now we show that the output of \fvs{} is \no if and only if  $\disjxy =0$.

From the construction, $(u_{a},w), (u_{b},w) \in E(G)$. If $\disjxy=0$,
 there exists $i \in [n]$ such that $x_i=y_i=1$. This implies the edges $(u_{a},v_{i})$  and $(u_{b},v_{i})$ are present in $G$. So, the cycle $\cC(u_{a},v_{i},u_{b},w)$ is present in $G$, that is, the output of \fvs is NO.
 
Conversely, if the output of \fvs is NO, there exists a cycle in $G$. From the construction, the cycle must be 
 $\cC(u_{a},$ $v_{i},u_{b},w)$ for some $i \in [n]$. As the edges $(u_{a},v_{i})$ and $(u_{b},v_{i})$ are present in $G$, $x_i=y_i=1$, that is, $\disjxy=0$. 

   Now by Propositions~\ref{pre:cc} and \ref{pre:cc2sc}(ii), we obtain that {\sc Feedback Vertex Set} parameterized by vertex cover size $K$ is $(\va, n/p,p)$-hard even if $\Delta_{av}(G)=\Oh(1)$, and when $k=0$.
\end{proof}
In each of the above three cases, we can make the reduction work for any $k$, by adding $k$ many vertex disjoint cycles of length $4$, i.e. $C_4$'s, to $G$. In Theorem~\ref{theo:lowerbounds1} (III), the vertex cover must be bounded. In the given reduction for Theorem~\ref{theo:lowerbounds1} (III), the vertex cover of the constructed graph is at most $2$.  Note that by the addition of $k$ many edge disjoint $C_4$'s, the vertex cover of the constructed graph in the modified reduction is at most $2k+2$, and is therefore still a parameter independent of the input instance size. 

This completes the proof of the Theorem~\ref{theo:lowerbounds1} with respect to \fvs.
 
 If the graph constructed in the reduction, in any of the above three cases for {\sc Feedback Vertex Set}, contains a cycle, then it is of even length. Otherwise, the graph is cycle free. Hence, the proof of this Theorem with respect to \ect is same as the proof for \fvs.
 
 Similarly, a slight modification can be made to the constructed graph, in all three of the above cases, such that a cycle in the graph is of odd length if a cycle exists. Thereby, the proof of this Theorem with respect to \Oct also is very similar to the proof for \fvs. 
\end{proof}

\begin{proof}[{\bf Proof of Theorem~\ref{theo:lowerbounds2}}]
We first show the hardness results of \tdel for $k=0$ in all three cases.

\begin{figure}[h!]
  \centering
  \includegraphics[width=8cm, height=6cm,keepaspectratio]{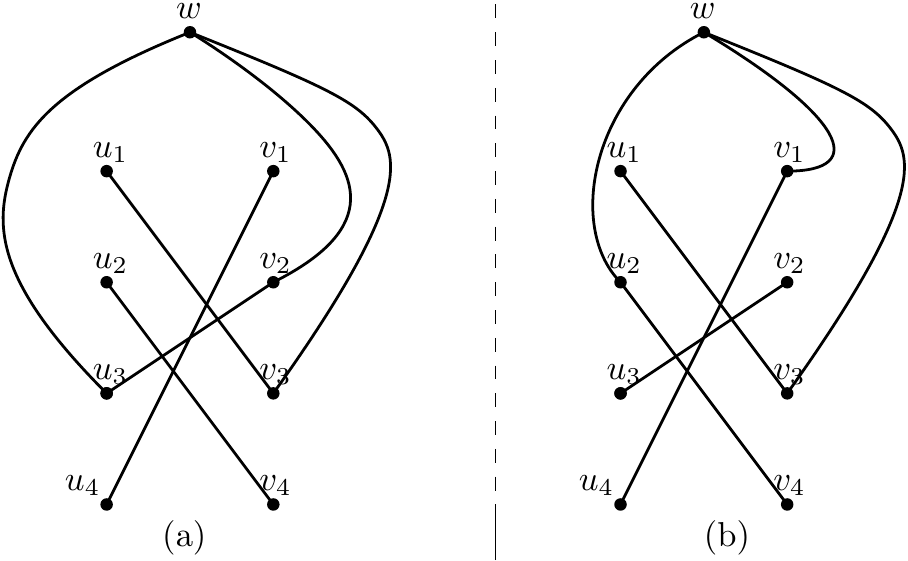}
  \caption{Illustration of Proof of Theorem~\ref{theo:lowerbounds2}~(I). Consider $n=4$. Let $\pi:[4]\rightarrow [4]$ such that $\pi(1)=3, \pi(2)=4, \pi(3)=2$, and $\pi(4)=1$. So the concatenated bit string is $11001001$. In (a), $j=5$, $\Phi(j)=(\psi,\gamma)=(3,1)$, $\permxy=1$ and $G$ contains a triangle. In (b), $j=4$, $\Phi(j)=(\psi,\gamma)=(2,2)$, $\permxy=0$, and $G$ does not contain any triangle.}
  \label{fig:tdel1}
\end{figure}
\begin{proof}[Proof of Theorem~\ref{theo:lowerbounds2}~(I)] We give a reduction from \perm to \tdel when the solution size parameter $k=0$. Let $\cA$ be a one pass streaming algorithm that solves \tdel in \va model, such that $\Delta_{av}(G)=\Oh(1)$, and the space used is $o(n \log n)$. 
\remove{Now we give a protocol that solves \perm using $o(n \log n)$ space. }
Let $G$ be a graph with $2n+1$ vertices $u_1,\ldots,u_n, v_1,\ldots,v_n, w$. Let $\pi$ be the input of Alice for \perm. See Figure~\ref{fig:tdel1} for an illustration.

{\bf Alice's input to $\cA$:} Alice inputs the graph $G$ by exposing the vertices $u_1,\ldots,u_n, v_1,\ldots,$ $v_n$, sequentially. (i) While exposing the vertex $u_i$, Alice does not give any edge; (ii) while exposing the vertex $v_i$, Alice gives the edges $(v_{\pi(i)},u_i)$ as an input to the stream of $\cA$. 

After the exposure of $u_1,\ldots,u_n, v_1,\ldots,$ $v_n$ as per the \va model, Alice sends the current memory state of $\cA$, i.e. the sketch generated by $\cA$, to Bob. Let $j \in [n \log n]$ be the input of Bob and let $(\psi,\gamma) =\Phi(j)$. 

{\bf Bob's input to $\cA$:} Bob exposes only the vertex $w$. Bob gives the edge $(w,u_\psi)$, and the edge $(w,v_i)$ if and only if $\bit(i,\gamma)=1$, as input to the stream of $\cA$.

 From the construction, note that $\Delta_{av}(G)=\Oh(1)$. Recall that $k=0$. Now we show that, the output of \tdel{} is \no if and only if $\permxy=1$.

From the construction, the edges $(u_\psi,v_{\pi(\psi)})$ and $(w,u_\psi)$ are present in $G$. If $\permxy=1$, then $(v_{\pi(\psi)},w) \in E(G)$. So, there exists a triangle in $G$, that is, the output of \tdel is NO.

On the other hand, if the output of \tdel is NO, then there exists a triangle in $G$. From the construction, the triangle is formed  
with the vertices $u_\psi,v_{\pi(\psi)}$ and $w$. As $(v_{\pi(\psi)},w) \in E(G)$, the $\gamma$-{{th}} bit of $\pi(\psi)$ is $1$, that is, $\permxy=1$.

   Now by Propositions~\ref{pre:cc} and \ref{pre:cc2sc}(iii), we obtain that \tdel is $(\va, n\log n)$-hard even if $\Delta_{av}(G)=\Oh(1)$, and when $k=0$.
\end{proof}
\begin{figure}[h!]
  \centering
  \includegraphics[width=0.7\linewidth]{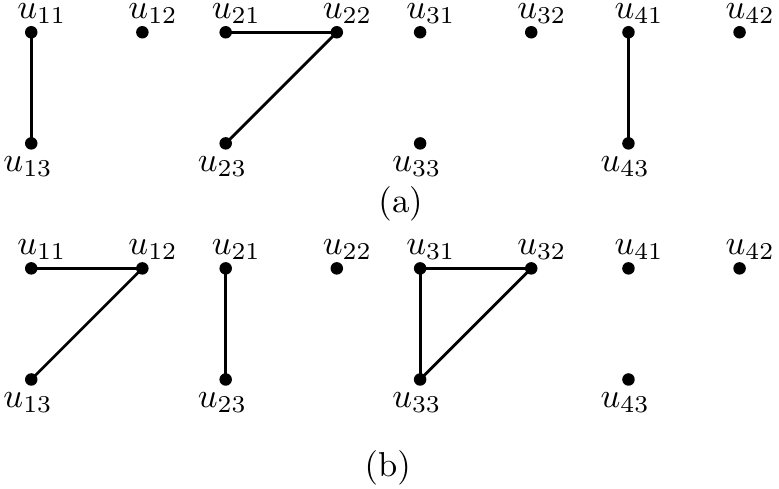}
  \caption{Illustration of Proof of Theorem~\ref{theo:lowerbounds2}~(II). Consider $n=4$. In (a), $\bx=1001$ and $\by=0100$, that is, $\disjxy=1$, and $G$ does not contain any triangle. In (b), $\bx=0110$ and $\by=1010$, that is, $\disjxy=0$, and $G$ contains a triangle.}
  \label{fig:tdel2}
\end{figure}
\begin{proof}[Proof of Theorem~\ref{theo:lowerbounds2}~(II)]  We give a reduction from \disj to \tdel when the solution size parameter $k=0$. Let $\cA$ be a one pass streaming algorithm that solves \tdel in \va model, such that $\Delta(G) =\Oh(1)$, and the space used is $o(n)$. Let $G$ be a graph with $3n$ vertices $u_{11},u_{12},u_{13}, \ldots, u_{n1},u_{n2},u_{n3}$. Let ${\bf x,y}$ be the input of Alice and Bob for \disj. See Figure~\ref{fig:tdel2} for an illustration.

{\bf Alice's input to $\cA$:} Alice inputs the graph $G$ first by exposing the vertices $u_{11},u_{12},u_{21},u_{22} \ldots, 
u_{n1},$ $u_{n2}$, sequentially. (i) While exposing $u_{i1}$, Alice does not give any edge; (ii) while exposing $u_{i2}$, 
Alice gives the edge $(u_{i2},u_{i1})$, if and only if $x_i=1$, as inputs to $\cA$. 

After the exposure of $u_{11},u_{12},u_{21},u_{22} \ldots, 
u_{n1},$ $u_{n2}$ as per the \va model, Alice sends current memory state of $\cA$, i.e. the sketch generated by $\cA$, to Bob. 

{\bf Bob's input to $\cA$: }Bob exposes the vertices $u_{13}, \ldots, u_{n3}$, sequentially. While exposing $u_{i3}$, Bob gives the edges $(u_{i3},u_{i1})$ and $(u_{i3},u_{i2})$ as two inputs to $\cA$ if and only if $ y_i=1$. 

 From the construction, note that $\Delta(G) \leq 2$. Recall that $k=0$. Now we show that the output of \tdel{} is \no if and only if $\disjxy =0$.

If $\disjxy=0$, there exists $i \in [n]$ such that $x_i=y_i=1$. From the construction, the edges $(u_{i2},u_{i1})$, $(u_{i3},u_{i1})$ and $(u_{i3},u_{i2})$ are present in $G$. So, there exists a triangle in $G$, that is, the output of \tdel is NO.

Conversely, if the output of \tdel is NO, there exists a triangle in $G$. From the construction, the triangle is 
$(u_{i1},u_{i2},u_{i3})$ for some $i \in [n]$. As the edge $(u_{i2},u_{i1}) \in E(G)$, $x_i=1$; and  as the edges $(u_{i3},u_{i1})$ and $(u_{i3},u_{i2})$ are in $G$, $y_i=1$. So, $\disjxy=0$.  

   Now by Propositions~\ref{pre:cc} and \ref{pre:cc2sc}(ii), we obtain that \tdel is $(\va, n/p,p)$-hard even if $\Delta(G)=\Oh(1)$, and when $k=0$.

\end{proof}
\begin{figure}[h!]
  \centering
  \includegraphics[width=0.7\linewidth]{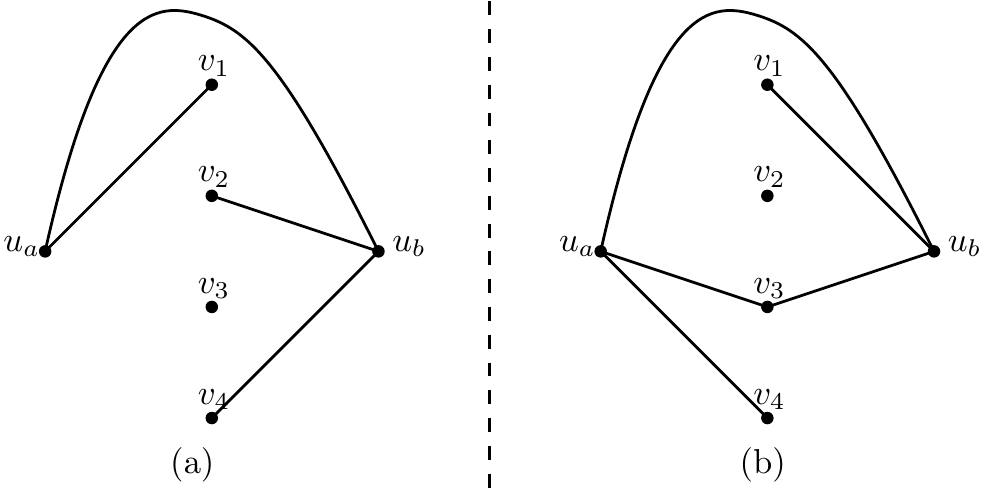}
  \caption{Illustration of Proof of Theorem~\ref{theo:lowerbounds2}~(III). Consider $n=4$. In (a), $\bx=1000$ and $\by=0101$, that is, $\disjxy=1$, and $G$ does not contain any triangle. In (b), $\bx=0011$ and $\by=1010$, that is, $\disjxy=0$, and $G$ contains a triangle.}
  \label{fig:tdel3}
\end{figure}
\begin{proof}[Proof of Theorem~\ref{theo:lowerbounds2}~(III)]  We give a reduction from \disj to \tdel parameterized by vertex cover size $K$, where $\cA$ is a one pass streaming algorithm that solves \tdel parameterized by $K$ in \va model such that $\Delta_{av}(G)=\Oh(1)$, and the space used is $o(n)$. Let $G$ be a graph with $n+2$ vertices $u_a,v_1,\ldots,v_n,u_b$. Let ${\bf x,y}$ be the input of Alice and Bob for \disj. See Figure~\ref{fig:tdel3} for an illustration.

{\bf Alice's input to $\cA$:} Alice inputs the graph $G$ first by exposing the vertices $u_a, v_1,\ldots,v_n$ sequentially. (i) While exposing $u_a$, Alice does not give any edge; (ii) while exposing $v_i$, Alice gives the edge $(v_i,u_a)$ as input to $\cA$ if and only if $x_i=1$. 

After the exposure of $u_a, v_1,\ldots,v_n$ as per the \va model, Alice sends current memory state of $\cA$, i.e. the sketch generated by $\cA$, to Bob. 

{\bf Bob's input to $\cA$:} Bob  exposes $u_b$ only. Bob gives the edge $(u_b,u_a)$ unconditionally, and an edge $(u_b,v_i)$ as input to $\cA$ if and only if $y_i=1$.

 From the construction, observe that $\vc(G)\leq 2 \leq K$ and $\Delta_{av}(G) =\Oh(1)$. Recall that $k=0$. Now we show that the output of \tdel{} is \no if and only if  $\disjxy =0$.

Observe that $(u_a,u_b) \in E(G)$. If $\disjxy=0$, there exists an $i \in [n]$ such that $x_i=y_i=1$. 
From the construction, the edges $(v_i,u_a)$ and $(u_b,v_i)$ are present in $G$. So, $G$ contains the triangle with vertices $u_a,u_b$ and $w$, i.e., the output of \tdel is NO.

On the other hand, if the output of \tdel is NO, there exists a triangle in $G$. From the construction, the triangle is formed with the vertices $u_a,u_b$ and $v_i$. As $(v_i,u_a)\in E(G)$ implies $x_i=1$, and  $(v_i,u_a) \in E(G)$ implies $y_i=1$. So, $\disjxy=0$.

Now by Propositions~\ref{pre:cc} and \ref{pre:cc2sc}(ii), we obtain that \tdel parameterized by vertex cover size $K$ is $(\va, n/p,p)$-hard even if $\Delta_{av}(G)=\Oh(1)$, and when $k=0$.
\end{proof}

In each of the above cases, we can make the reductions work for any $k$, by adding $k$ many vertex disjoint triangles to $G$. In Theorem~\ref{theo:lowerbounds2} (III), the vertex cover must be bounded.  In the given reduction for Theorem~\ref{theo:lowerbounds2} (III), the vertex cover of the constructed graph is at most $2$.  Note that by the addition of $k$ many edge disjoint $C_4$'s, the vertex cover of the constructed graph in the modified reduction is at most $2k+2$, and is therefore still a parameter independent of the input instance size. 

Hence, we are done with the proof of the Theorem~\ref{theo:lowerbounds2}.
\end{proof}

\begin{figure}[h!]
  \centering
  \includegraphics[width=0.8\linewidth]{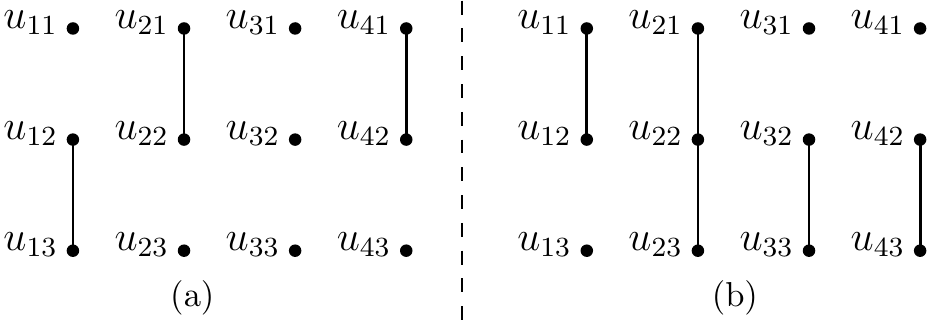}
  \caption{Illustration of Proof of Theorem~\ref{theo:lowerbounds3}. Consider $n=4$. In (a), $\bx=0101$ and $\by=1000$, that is, $\disjxy=1$, and $G$ does not have any induced $P_3$. In (b), $\bx=1100$ and $\by=0112$, that is, $\disjxy=0$, and $G$ contains an induced $P_3$.}
  \label{fig:cvd}
\end{figure}

\begin{proof}[{\bf Proof of Theorem~\ref{theo:lowerbounds3}}]
We give a reduction from \disj to \cvd for solution size parameter $k=0$. Let $\cA$ be a one pass streaming algorithm that solves \cvd in \va model, such that $\Delta(G) =\Oh(1)$, and the space used is $o(n)$. Consider a graph $G$ with $3n$ vertices $u_{11},u_{12},u_{13}, \ldots, u_{n1},u_{n2},$ $u_{n3}$. Let 
${\bf x,y}$ be the input of Alice and Bob for \disj. See Figure~\ref{fig:cvd} for an illustration.

{\bf Alice's input to $\cA$:} Alice inputs the graph $G$ by exposeing the vertices $u_{11},u_{12},u_{21},u_{22} \ldots, 
u_{n1}, $ $u_{n2}$, sequentially. (i) While exposing $u_{i1}$, Alice does not give any edge; (ii) while exposing $u_{i2}$, 
Alice gives the edge $(u_{i2},u_{i1})$ as input to $\cA$ if and only if $x_i=1$. 

After the exposure of $u_{11},u_{12},u_{21},u_{22} \ldots, 
u_{n1}, $ $u_{n2}$ as per the \va model, Alice sends current memory state of $\cA$, i.e. the sketch generated by $\cA$, to Bob. 

{\bf Bob's input to $\cA$:} Bob exposes the vertices $u_{13}, \ldots, u_{n3}$, sequentially. While exposing $u_{i3}$, Bob gives the edges $(u_{i3},u_{i2})$ as an input to $\cA$ if and only if $ y_i=1$. 

 From the construction, note that $\Delta(G) \leq 2$. Observe that, there exists a $P_3$ in $G$ if and only if there exists an $i \in [n]$ such that $x_i=y_i=1$. Hence,
 the output of \cvd{} is \no if and only if $\disjxy =0$. 
 
 Now by Propositions~\ref{pre:cc} and \ref{pre:cc2sc}(ii), we obtain that \cvd is $(\va, n/p,p)$-hard even if $\Delta(G)=\Oh(1)$, and when $k=0$.

We can make the reduction work for any $k$, by adding $k$ many vertex disjoint $P_3$'s to $G$.
\end{proof}

\section{Conclusion}
In this paper, we initiate the study of parameterized streaming complexity with structural parameters for graph deletion problems. Our study also compared the parameterized streaming complexity of several graph deletion problems in the different streaming models. 
 In future, we wish to investigate why such a classification exists for seemingly similar graph deletion problems, and conduct a systematic study of other graph deletion problems as well.   

\bibliographystyle{alpha}
\bibliography{reference}

\newpage
\appendix

\section{Problem Definitions}\label{sec:probdefi}
\noindent In this Section we define the following problems formally.
\remove{We study the parameterized versions of the following optimization problems. The parameters we consider in this paper are (i) the solution size $k$ as parameter and (ii) the size $K$ of the vertex cover of the input graph $G$ as a structural parameter.}

\defproblem{\subdel}{A graph $G$, a family $\cF$ of connected graphs, and a non-negative integer $k$.}{Does there exist a set $X \subset V(G)$ of 
$k$ vertices such that $G \setminus X$ does not contain any graph from $\cF$ as a subgraph?}

\defproblem{\minordel}{A graph $G$, a family $\cF$ of connected graphs, and a non-negative integer $k$.}{Does there exist a set $X \subset V(G)$ of 
$k$ vertices such that $G \setminus X$ does not contain any graph from $\cF$ as a minor?}

\defproblem{\fvs}{A graph $G$ and a non-negative integer $k$.}{Does there exist a set $X \subset V(G)$ of 
$k$ vertices such that $G \setminus X$ does not contain any cycle?}

\defproblem{\ect}{A graph $G$ and a non-negative integer $k$.}{Does there exist a set $X \subset V(G)$ of 
$k$ vertices such that $G \setminus X$ does not contain any cycle of even length?}

\defproblem{\Oct}{A graph $G$ and a non-negative integer $k$.}{Does there exist a set $X \subset V(G)$ of 
$k$ vertices such that $G \setminus X$ does not contain any cycle of odd length, i.e., $G \setminus X$ is bipartite?}

\defproblem{\tdel}{A graph $G$ and a non-negative integer $k$.}{Does there exist a set $X \subset V(G)$ of 
$k$ vertices such that $G \setminus X$ does not contain any triangle?}

\defproblem{\cvd}{A graph $G$ and a non-negative integer $k$.}{Does there exist a set $X \subset V(G)$ of 
$k$ vertices such that $G \setminus X$ is a cluster graph, i.e., $G \setminus X$ does not contain any induced $P_3$?}

\defproblem{\cn}{A graph $G$ with $\vc(G) \leq K$, degree parameter $d \leq K$ and common neighbor parameter $\ell$.}{A common neighbor subgraph of $G$.}
\remove{Notice that for all the above problems $K \geq k$. \remove{of the vertex cover of the input graph $G$ is always at least the size $k$ of the solution.} Therefore, our motivation for using $K$ as a parameter was to try and see if problems that are hard in the streaming setting under the natural parameter $k$ become streamable under the larger parameter $K$. Interestingly, the parameter $K$ has different effects on the above problems, in the different streaming models.}

\end{document}